\pdfoutput=1
\documentclass{article}

\usepackage{arxiv}
\usepackage{natbib}
\usepackage[utf8]{inputenc} 
\usepackage[T1]{fontenc}    
\usepackage{hyperref}       
\usepackage{url}            
\usepackage{booktabs}       
\usepackage{amsfonts}       
\usepackage{nicefrac}       
\usepackage{microtype}      
\usepackage{lipsum}
\usepackage{amsmath}
\usepackage{graphicx}
\usepackage{subfigure}
\usepackage{amsfonts}
\usepackage[toc,page]{appendix}
\IfFileExists{upquote.sty}{\usepackage{upquote}}{}
\setcitestyle{square}
\title{\emph{SQUAREM}: An \textbf{R} Package for Off-the-Shelf Acceleration of EM, MM and Other EM-like Monotone Algorithms}

\author{
  Yu Du\\
  Research Scientist\\
  Global Statistical Sciences\\
  Eli Lilly and Company\\
  Indianapolis, IN 46285 \\
  \texttt{du\_yu@lilly.com}\\
   \And
  Ravi Varadhan\thanks{Corresponding Author}\\
  Associate Professor\\
  Department of Oncology \\
  Johns Hopkins University\\
  Baltimore, MD 21287\\
  \texttt{ravi.varadhan@jhu.edu}\\
}

\begin{document}
\maketitle

\begin{abstract}
We discuss \textbf{R} package \emph{SQUAREM} for accelerating iterative algorithms which exhibit slow, monotone convergence.  These include the well-known expectation-maximization algorithm, majorize-minimize (MM), and other EM-like algorithms such as expectation conditional maximization, and generalized EM algorithms. 
We demonstrate the simplicity, generality, and power of \emph{SQUAREM} through a wide array of applications of EM/MM problems,
including binary Poisson mixture, factor analysis, interval censoring, 
genetics admixture, and logistic regression maximum likelihood estimation (an MM problem). We show that
\emph{SQUAREM} is easy to apply, and can accelerate any fixed-point, smooth, contraction mapping 
with linear convergence rate. Squared iterative scheme (Squarem) algorithm provides significant speed-up of EM-like algorithms.  The margin of the advantage for Squarem is especially huge for high-dimensional problems or when EM step is relatively time-consuming to evaluate.  Squarem can be used off-the-shelf since there is no need for the user to tweak any control parameters to optimize performance.  
Given its remarkable ease of use, Squarem may be considered as a default accelerator for slowly converging EM-like algorithms.
All the comparisons of CPU computing time in the paper are made on a quad-core 2.3 GHz Intel Core i7 Mac computer. 
\textbf{R} Package \emph{SQUAREM} can be downloaded at \url{https://cran.r-project.org/web/packages/SQUAREM/index.html}.
\end{abstract}


\section{Introduction}\label{introduction}
The \textbf{R} package \emph{SQUAREM} provides convergence acceleration techniques for speeding-up slow, monotone iterative algorithms. These include the well-known expectation-maximization (EM) algorithm\citep{dempster1977maximum}, majorize-minimize (MM)\citep{lange2000optimization}, and other algorithmic variants such as expectation-conditional maximization (ECM)\citep{meng1993maximum}, expectation-conditional maximization or either (ECME)\citep{liu1998maximum}, among others. \cite{dempster1977maximum} refer to such variants as ``generalized EM (GEM)" when M step is only partially implemented. In this paper, we term these ``EM-like algorithms'', because they all have a contractive fixed-point mapping with linear rate of convergence, like EM. For the definition of linear rate of convergence and contractive mapping, please refer to \cite[Chapter~5]{ortega1970iterativebook}. All of these algorithms are essentially based on the idea that a relatively difficult optimization problem can be converted to a much simpler iterative algorithm with guaranteed, albeit slow, convergence.  A visual imagery is apt here: instead of embarking upon a direct and treacherously steep climb, we approach the summit through a winding, gradually ascending path. Interestingly, this idea has become very attractive now with the advent of big data revolution and high-dimensional applications, where solving the original optimization problem is either impossible or prohibitively expensive. There is a nice analogy to this in numerical linear algebra for solving large-scale linear system of equations. Indirect, iterative techniques (e.g., Gauss-Seidel) for solving linear systems were considered to be too slow and impractical, and only of pedagogical interest.  Instead, the attention of the research community was focused on direct methods such as the various decomposition and factorization methods (LU, QR, SVD).  But, such direct techniques are ill-suited to solve the modern day, large-scale linear systems with millions of equations.  Therefore, clever adaptations of indirect iterative methods are emerging as the methods of choice (e.g., conjugate-gradient)\citep{censor1997parallel,saad2003iterative}.  Similarly, for the estimation of statistical models in large, high-dimensional modern applications, EM-like algorithms are becoming indispensable tools in the arsenal of computational scientists \citep[etc.]{patro2014sailfish, shiraishi2015simple, raj2015mscentipede, chiou2017semiparametric}. 

\par EM-like algorithms are characterized by two essential features: reliable, monotone convergence, and slow, linear rate of convergence.  Therefore, any strategy that can accelerate the rate of convergence of these algorithms, without compromising on their reliability and ease of use, will be of huge help. \cite{Zhou2011} remarked that ``In many statistical problems, maximum likelihood estimation by an EM or MM algorithm suffers from excruciatingly slow convergence. This tendency limits the application of these algorithms to modern high-dimensional problems in data mining, genomics, and imaging. Unfortunately, most existing acceleration techniques are ill-suited to complicated models involving large numbers of parameters. The squared iterative methods (SQUAREM) recently proposed by Varadhan and Roland constitute one notable exception.'' The goal of this paper is to demonstrate the utility of this ``notable exception", Squarem, proposed by \cite{varadhan2008simple}, which is available in the \textbf{R} package \emph{SQUAREM} \citep{varadhan2016squarem}. 

\par The main aim of \emph{SQUAREM} is to facilitate the development of computationally efficient new statistical models.  In particular, our package provides acceleration schemes which can speed up the estimation of the statistical models, where the model parameters are estimated with monotone, EM-like algorithms. Acceleration of these estimation algorithms can be readily achieved using the function \verb+squarem()+.  Here we demonstrate the simplicity, generality, and power of \emph{SQUAREM} through a wide array of applications of EM/MM problems in R\citep{renv2015}, illustrating how easy it is to use Squarem to derive efficient solutions. However, it should be recognized that there is no foolproof numerical algorithm; in poorly identified problems, where even the EM algorithm can fail, Squarem is not guaranteed to work.

\section{Squared iterative method}\label{squared-iterative-method}
Suppose we have observed data $y = (y_1,\hdots,y_N)^\top $ that comes from a probability density function $g(y;\theta)$ where $\theta \in \Omega \subset \mathbb{R}^p$ is the parameter of interest. We are often interested in computing the MLE (maximum likelihood estimates) of $\theta$, denoted by $\theta^{\star}.$ EM algorithm is a popular technique for computing MLE, which consists of two steps, E-step and M-step\citep{dempster1977maximum}. EM algorithm is natural when there is a missing data $z$ component in the probability model, which when known greatly simplifies the estimation of $\theta^{\star}.$ Let us use $x = \lbrace y, z \rbrace$ to denote the complete data. EM algorithm then becomes:

\begin{itemize}
\item E-step

A $Q$ function is constructed such that $$Q(\theta; \theta_n) = \int L_c(\theta;x)f(z;y,\theta_n)dz, \quad n=0,1,\hdots,$$ where $n$ refers to the $n^{th}$ iteration of the algorithm, $L_c(\theta;x)$ is the complete data log-likelihood,  and $f(z;y,\theta_n)$ is the conditional density function of missing data $z$ given observed data $y$. Thus, the $Q$ function computes the expected value of complete data log-likelihood given the current estimates of parameter values and observed data.

\item M-step

M-step maximizes the $Q$ function in E-step over $\theta \in \Omega \subset \mathbb{R}^p$ to iteratively compute the next, $(n+1)^{th}$ iteration of parameter values, $\theta_{n+1}$, such that $$\theta_{n+1} = \arg\!\max Q(\theta; \theta_n), \quad n=0,1,\hdots,.$$

\end{itemize}

The EM algorithm therefore defines a fixed-point mapping $F$ such that $F:\Omega \subset \mathbb{R}^p \mapsto \Omega$ and $$\theta_{n+1} = F(\theta_n), \quad n=0,1,\hdots.$$ Two convergence criteria can be applied and are both satisfied by the EM algorithm : 1) for parameter estimates $\theta_n$, as $n \xrightarrow{}{} \infty$, $||\theta_n - \theta^{\star}|| \xrightarrow{}{} 0$ ($||.||$ is the Euclidean Norm); 2) the convergence is defined by the sequence of the likelihood function of the parameter estimates, $L(\theta_n)$, such that $|L(\theta_n) - L(\theta^{\star})| \xrightarrow{}{} 0$, as $n \xrightarrow{}{} \infty$. The EM algorithm guarantees to produce monotone convergence such that $L(\theta_{n+1}) \geq L(\theta_n)$. By Taylor's theorem under regularity conditions, expand $F(\theta_n)$ around $\theta^{\star}$:
$$\theta_{n+1} - \theta^{\star} = J(\theta^{\star})(\theta_n - \theta^{\star}) + o(||\theta_n - \theta^{\star}||)$$ where $J(\theta^{\star})$ is the Jacobian matrix of $F$ evaluated at $\theta^{\star}.$ \cite{dempster1977maximum} showed that $J(\theta^{\star})$, the Jacobian matrix, measures the fraction of missing information. Under weak regularity conditions, the eigenvalues of $J(\theta^{\star})$ lie on $[0, 1).$ Thus, the largest eigenvalue of $J(\theta^{\star})$ governs the rate of convergence for EM. The closer this is to unity, indicating a large fraction of missing information, the slower EM converges. 

Motivated by the Cauchy-Barzilai-Borwein(CBB) method \citep{raydan2002relaxed}, \cite{roland2005new} and \cite{varadhan2008simple} constructed Squarem by defining the following recursive error relation: $$e_{n+1} = [I - \alpha_n(J - I)]^2e_n,$$ where $e_n = \theta_n - \theta^{\star}$, $I$ is the identity matrix and $\alpha_n$ is the steplength that takes into account the larger eigenvalues of $J(\theta^{\star})$. The pseudocode for Squarem algorithm is listed in Table~\ref{sqpseudo} \citep{varadhan2008simple}, which demonstrates the remarkable simplicity of the proposed algorithm.

\begin{table}[ht]
\centering
\begin{tabular}{l}
\hline
Input: $F$, $L$, $\theta_0$, $\eta \geq 0$ \\
 While not converged \\
 \hline
 $\theta_{1} = F(\theta_0)$ \\
 $\theta_{2} = F(\theta_{1})$\\
 $r = \theta_{1} - \theta_0$\\
 $v = (\theta_{2}- \theta_{1}) - r$\\
 Compute steplength $\alpha$ \\
 $\theta_{sq} = \theta_0 - 2\alpha r + \alpha^2 v$\\
 If $L(\theta_{sq}) > L(\theta_2) - \eta$, set $\theta' = \theta_{sq}$; else  $\theta' =\theta_2$ \\
 $\theta_0 = F(\theta')$, stabilization step (done only if $\theta' = \theta_{sq}$) \\
  \hline
\end{tabular}
\caption{Pseudocode for Squarem.}
\label{sqpseudo}
\end{table}

There are three choices for $\alpha$, the steplength as described in \cite{varadhan2008simple}.  It is our experience that $\alpha = -\frac{||r||}{||v||}$ generally works the best, and hence it is the default steplength used in \emph{SQUAREM}. \cite{varadhan2008simple} also showed global convergence of Squarem algorithm, i.e., Squarem can converge to a stationary point from any starting value in the parameter space, or at least, in a large part of it by modifying steplength to ensure monotonicity. Note that when steplength is equal to $-1$, a Squarem evaluation is the same as two EM updates.  Thus, each iteration of Squarem involves 2 or 3 evaluations of EM. Hence, when we compare EM to Squarem, we use the number of EM steps rather than number of iterations.  Apart from the EM steps, there is minimal cost in computing the Squarem parameter updates, including the computation of the value of likelihood functions. In addition to the convergence criteria provided earlier, we give a definition of convergence acceleration as follows: suppose $\{\theta_n\}$ is the sequence of estimates produced by Algorithm 1, while $\{\theta_n'\}$ is that given by Algorithm 2, then we say that Algorithm 2 accelerates Algorithm 1 if $\frac{||\theta_n' - \theta^{\star}||}{||\theta_n - \theta^{\star}||} \xrightarrow{}{} 0$ as $n \xrightarrow{}{} \infty$.

\section[Description of R package SQUAREM]{Description of \textbf{R} package \emph{SQUAREM}}\label{descsq}
\textbf{R} Package \emph{SQUAREM} is available on CRAN. It can be downloaded via \url{https://cran.r-project.org/web/packages/SQUAREM/index.html}. \emph{SQUAREM} works for any smooth, contraction mapping with a linear convergence rate (e.g., EM-like algorithms). 
We describe below the two main functions, \verb+squarem()+ and \verb+fpiter()+. Obviously, \verb+squarem()+ is the featured function in the package. 
\begin{itemize}
\item \verb+squarem()+, for squared iterative scheme

\verb+squarem()+ is a function to accelerate any smooth, contractive, fixed-point iteration algorithm including EM/MM and other EM-like algorithms. The main arguments include \verb+par+, \verb+fixptfn+, \verb+objfn+ and \verb+control+. \verb+par+ denotes the starting value of parameters.  The argument \verb+fixptfn+ defines a function $F$ representing the fixed-point iteration: $\theta_{k+1} = F(\theta_{k})$. \verb+fixptfn+ encodes a single step of any EM-like algorithm.

\begin{verbatim}
R> fixptfn <- function(par, data, ...) {
+    pnew <- F(par, data, ...)
+    return(pnew)
+  }    
\end{verbatim}

\verb+objfn+ is the objective function we want to minimize.  In the case of EM-like algorithms, it would be the negative log-likelihood function of data.  It is not essential to supply the objective function in order for Squarem to work, but its provision guarantees global convergence. \verb+control+ specifies a list of algorithm options including \verb+maxiter+, maximum number of iterations, and \verb+tol+, tolerance, among others. If $||F(\theta_k) - \theta_k|| \leq \text{tol}$, the algorithm shall declare convergence at the $(k+1)^{th}$ iteration ($||.||$ shows the Euclidean Norm). Under regularity conditions given by \cite{jeffwu}, the satisfaction of the convergence does imply a local optimum. There are 3 important control parameters in \verb+squarem()+, namely, \verb+K+, \verb+method+ and \verb+objfn.inc+. \verb+method+ specifies the choice of steplength and \verb+K+ specifies the order of the squared iterative scheme.  The default values \verb+method = 3+ and \verb+K = 1+ generally work well. \verb+objfn.inc+ guides the monotonicity of the objective function.  Setting \verb+objfn.inc = 0+ ensures strict monotonicity, while \verb+objfn.inc = Inf+ results in an unguarded acceleration scheme, where the objective function is not evaluated at all. The default is \verb+objfn.inc = 1+, which results in a nearly-monotone acceleration scheme.  We can also set this to equal the average log-likelihood i.e., log-likelihood per individual sample.

 To summarize, the default usage of \verb+squarem()+ is such that 
 
 \verb+squarem(par, fixptfn, objfn, ..., control = list())+.

\item \verb+fpiter()+, for fixed-point iteration scheme

\verb+fpiter()+ is a function to implement the fixed-point iteration algorithm including EM, MM and other EM-like algorithms, without any acceleration. The main arguments include \verb+par+, \verb+fixptfn+, \verb+objfn+ and \verb+control+, which work the same way as in the \verb+squarem()+ except that there are no Squarem specific control parameters in the argument \verb+control+.


To summarize, the default usage of \verb+fpiter()+ is such that 

\verb+fpiter(par, fixptfn, objfn, ..., control = list())+.

\end{itemize}
In the next section, we demonstrate a detailed illustration of how to use \emph{SQUAREM}.

\section[How to apply SQUAREM acceleration]{How to apply \emph{SQUAREM} acceleration}\label{howto}
Imagine that an EM-like algorithm is used to estimate a model, with slow, linear rate of convergence. In order to speed up the algorithm using \textbf{R} Package \emph{SQUAREM}, there are two main steps to be prepared. 

\begin{itemize}
\item Step 1

Write the part from the used EM-like algorithm into an R function such that it does only one step of the EM-like algorithm. This function corresponds to the argument \verb+fixptfn+ in function \verb+squarem()+.

\item Step 2

Write an associated merit function to minimize, for example, the negative log likelihood function. This function corresponds to the argument \verb+objfn+ in function \verb+squarem()+.

\end{itemize}

There are also other arguments in function \verb+squarem()+ as specified in Section~\ref{descsq}, such as starting values, tolerance and maximum number of iterations, but the default choices often work well. Once we have these arguments ready, we can launch the function \verb+squarem()+ to do the acceleration, simple and easy. We now illustrate this usage of \textbf{R} Package \emph{SQUAREM} in detail with a simple example of mixture problem introduced below, which was also discussed in \cite{varadhan2008simple}. Here we revisit this example, mainly to illustrate how remarkably easy it is to apply the \verb+squarem()+ function, starting from an existing EM algorithm function.

In many studies, the study sample comes from a population which is a mix of two or more types of units, each with varying characteristics.  Finite mixture models are ideally suited to account for this kind of heterogeneity. A finite mixture model estimates parameters describing each subpopulation and their mixing probabilities. EM algorithm is a popular technique to compute the maximum likelihood estimates for mixture models, but is notorious for its slow convergence. Here, we use a two-component Poisson mixture to illustrate the usage and power of Squarem compared to the EM algorithm. We use the data of the number of deaths of women 80 years and older during the years 1910-1912 from \textit{The London Times} \citep{hasselblad1969estimation}. 

We use $p$ to denote the mixing probability and let $\mu_1$, $\mu_2$ be the mean of Poisson distribution from population 1 and 2, respectively. Let $i$ be the number of death, $i=0,1,\hdots,9$ and $n_i$ be the number of days when death $i$ occurred.

The real data is displayed in Table~\ref{poidata}, where death number $i=0,1,\hdots,9$.

\begin{table}[h!]
\centering
\begin{tabular}{llll}
\hline
 Death,$i$           & Frequency, $n_i$        & Death,$i$      & Frequency, $n_i$ \\
 \hline
 0& 162 &5&61 \\
 1& 267& 6&27 \\
 2& 271& 7&8 \\
 3& 185& 8&3 \\
 4& 111& 9&1 \\
 \hline
\end{tabular}
\caption{Data on deaths of women 80 years or older during 1910 to 1912 from \textit{The London Times}.}
\label{poidata}
\end{table}

\paragraph{EM algorithm}\label{em-algorithmpoi}
For derivation of EM step, see Appendix~\ref{poisem}.

The EM update is such that 
\begin{align*}
p^{(k+1)} & = \frac{\sum_i n_i p_i^{(k)}}{\sum_i n_i},\\
\mu_1^{(k+1)} & = \frac{\sum_iin_ip_i^{(k)}}{\sum_in_ip_i^{(k)}},\\
\mu_2^{(k+1)} & = \frac{\sum_iin_i(1-p_i^{(k)})}{\sum_in_i(1-p_i^{(k)})},
\end{align*} where $p^{(k+1)}, \mu_1^{(k+1)}, \mu_2^{(k+1)}$ are the $(k+1)^{th}$ iteration of estimates and $p_i^{(k)}$ is defined in Appendix~\ref{poisem}.

Next, we demonstrate how easy it is to set up Squarem acceleration of an EM-like algorithm. We implement the EM algorithm using function \verb+EM.poisson.mixture()+ below.

\begin{verbatim}
R> EM.poisson.mixture <- function(p, maxiter = 5000, tol = 1e-08, y) {
+    iter <- 1
+    conv <- FALSE
+    pnew <- rep(NA,3)
+    while(iter < maxiter) {
+      i <- 0 : (length(y) - 1)
+      zi <- p[1] * exp(-p[2]) * p[2]^i / 
+        (p[1] * exp(-p[2]) * p[2]^i + (1 -
+        p[1]) * exp(-p[3]) * p[3]^i)
+      pnew[1] <- sum(y * zi) / sum(y)
+      pnew[2] <- sum(y * i * zi) / sum(y * zi)
+      pnew[3] <- sum(y * i * (1 - zi)) / sum(y * (1 - zi))
+      res <- sqrt(crossprod(pnew - p))
+      p <- pnew
+      if(res < tol) {
+        conv <- TRUE
+        break
+      }
+      iter <- iter + 1
+    }
+    return(list(par = p, fpevals = iter, convergence = conv))
+  }
\end{verbatim}

In order to implement squared iterative scheme(Squarem) using function \verb+squarem()+, we extract the part in the above EM function that corresponds to one EM step and put it into a separate function \verb+poissmix.em()+.  This function corresponds to the argument \verb+fixptfn+ in the \verb+squarem()+ function. Cutting and pasting the code chunk from the above function, we create the function for \verb+fixptfn+ and complete step 1 in applying Squarem.

\begin{verbatim}
R> poissmix.em <- function(p,y) {
+    pnew <- rep(NA,3)
+    i <- 0 : (length(y) - 1)
+    zi <- p[1] * exp(-p[2]) * p[2]^i / 
+      (p[1] * exp(-p[2]) * p[2]^i + (1 -
+      p[1]) * exp(-p[3]) * p[3]^i)
+    pnew[1] <- sum(y * zi) / sum(y)
+    pnew[2] <- sum(y * i * zi) / sum(y * zi)
+    pnew[3] <- sum(y * i * (1 - zi)) / sum(y * (1 - zi))
+    p <- pnew
+    return(pnew)
+  }
\end{verbatim}

Step 2 is to write an associated merit function to minimize, in this case, the negative log likelihood function. The log likelihood of observed data $i,n_i$ is such that:
$$ll(p, \mu_1, \mu_2) = \sum_i n_i (\log{[pe^{-\mu_1}\mu_1^i/i! + (1-p)e^{-\mu_2}\mu_2^i/i!]}).$$

Therefore, the negative log likelihood is coded into function \verb+poissmix.loglik()+. This function corresponds to the argument \verb+objfn+ in function \verb+squarem()+.

\begin{verbatim}
R> poissmix.loglik <- function(p,y) {
+    i <- 0 : (length(y) - 1)
+    loglik <- y * log(p[1] * exp(-p[2]) * p[2]^i / exp(lgamma(i + 1)) + 
+      (1 - p[1]) * exp( - p[3]) * p[3]^i / exp(lgamma(i + 1)))
+    return(-sum(loglik))
+  }
\end{verbatim}

Now, we are all set to apply \verb+squarem()+. Let us then compare Squarem and the EM algorithm with starting value $(p, \mu_1, \mu_2) = (0.3, 1, 5)$ and tolerance $10^{-8}$, to see how remarkably easy it is to apply Squarem, off-the-shelf acceleration, once the EM algorithm has been implemented. 

\begin{verbatim}
R> library("SQUAREM")
R> poissmix.dat <- data.frame(death = 0 : 9, 
+    freq = c(162, 267, 271, 185, 111, 61, 27, 8, 3, 1))
R> y <- poissmix.dat$freq
R> p0 <- c(0.3,1,5)
R> system.time(f0 <- EM.poisson.mixture(p = p0, y = y))
\end{verbatim}

\begin{verbatim}
user system elapsed
0.036 0.005 0.040
\end{verbatim}

\begin{verbatim}
R> f0
\end{verbatim}

\begin{verbatim}
$par
[1] 0.3598864 1.2560968 2.6634056

$value.objfn
[1] 1989.946 

$fpevals 
[1] 2696

$convergence
[1] TRUE
\end{verbatim}

\begin{verbatim}
R> system.time(f1 <- fpiter(par = p0, fixptfn = poissmix.em, 
+    objfn = poissmix.loglik, control = list(tol = 1.e-08), y = y))
\end{verbatim}

\begin{verbatim}
user system elapsed
0.039 0.000 0.039
\end{verbatim}

\begin{verbatim}
R> f1
\end{verbatim}

\begin{verbatim}
$par
[1] 0.3598864 1.2560968 2.6634056

$value.objfn
[1] 1989.946

$fpevals
[1] 2696

$objfevals
[1] 0

$convergence
[1] TRUE
\end{verbatim}

\begin{verbatim}
R> system.time(f2 <- squarem(par = p0, fixptfn = poissmix.em, 
+    objfn = poissmix.loglik, control = list(tol = 1.e-08), y = y))
\end{verbatim}

\begin{verbatim}
user system elapsed
0.003 0.000 0.002
\end{verbatim}

\begin{verbatim}
R> f2
\end{verbatim}

\begin{verbatim}
$par
[1] 0.3598859 1.2560960 2.6634050

$value.objfn
[1] 1989.946

$iter
[1] 19

$fpevals
[1] 54

$objfevals
[1] 19

$convergence
[1] TRUE
\end{verbatim}

The output shows the equivalence of the standard EM loop function \verb+EM.poisson.mixture()+ and function \verb+fpiter()+ in \textbf{R} Package \emph{SQUAREM}, and a dramatic improvement for Squarem over the EM algorithm. Squarem outperforms EM for this case by a factor of 50 in terms of the number of EM evaluations and by a factor of 20 with regards to the CPU running time. From this point on, we will use function \verb+fpiter()+ to implement EM and other EM-like algorithms.

We conduct two algorithms for 5000 randomly generated starting values 
$$(p, \mu_1, \mu_2) = (U[0.05,0.95], U[0,20], U[0,20])$$ where $U[a,b]$ is a uniform random variable on the interval $[a,b]$. Table~\ref{poires2} displays the results. We provide the mean and 95\% confidence interval for the number of EM evaluations, fevals, as well as the CPU time (in seconds). In general, Squarem converges 13 times faster than EM with only 3.2\% of the number of EM evaluations that the EM algorithm takes.

\begin{table}[h!]
\centering
\begin{tabular}{lll}
\hline
                  & fevals           & CPU time(s)      \\
                  \hline
EM          & 3140(2510, 3182) & 0.039(0.031, 0.046) \\
Squarem     & 101(57,132)      & 0.003(0.002, 0.005)   \\
\hline
\end{tabular}
\caption{The convergence performance in terms of the number of EM evaluations, fevals and the CPU running time comparing EM to Squarem for Poisson mixture estimation with 5000 randomly generated starting values.}
\label{poires2}
\end{table}

\begin{figure}[h!]
\centering
\includegraphics[width=1\textwidth]{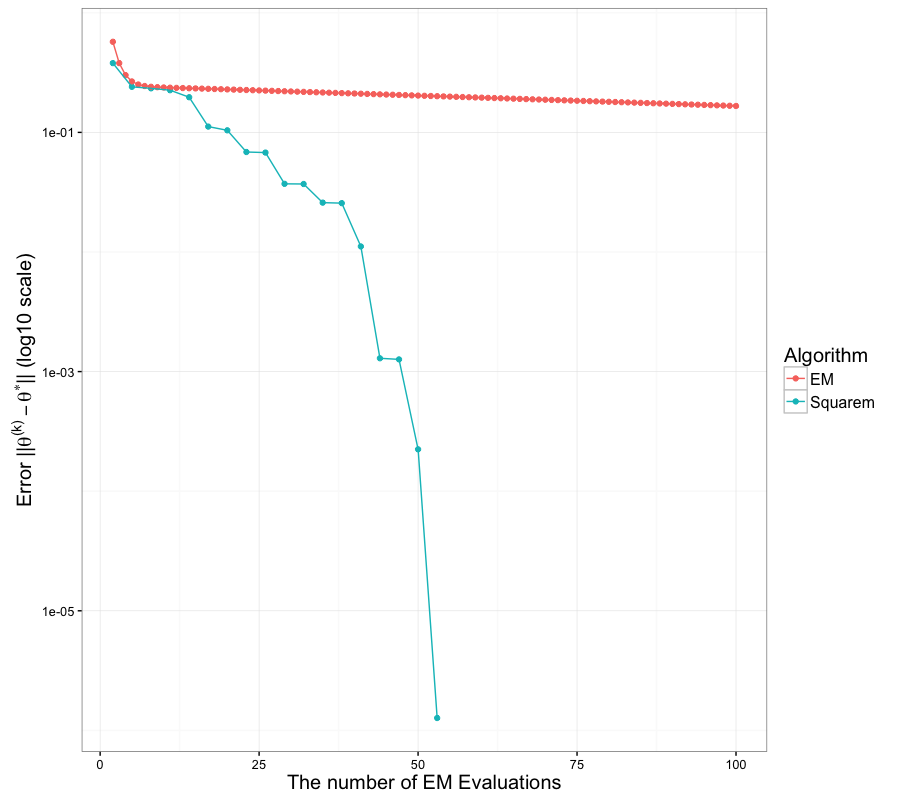}
\caption{The comparison of convergence behaviour between Squarem and EM.}
\label{poisres3}
\end{figure}

We also plot the error curve $||\theta^{(k)} - \theta^{*}||$ as a function of the number of EM evaluations $k$ in Figure~\ref{poisres3} using the starting value $(p, \mu_1, \mu_2) = (0.3, 1, 5)$, where $\theta^{(k)}$ is the $k^{th}$ iteration estimates $(p^{(k)}, \mu_1^{(k)}, \mu_2^{(k)})$.  The truth, $\theta^{*}$, is derived by running \verb+squarem()+ with a very small convergence tolerance, e.g., $10^{-18}$. Figure~\ref{poisres3} shows that the error for Squarem algorithm drops at a much faster rate than EM. Squarem converges in approximately 50 EM evaluations while the error is still quite large for EM after 100 iterations.  

In the next section, we continue to demonstrate the utility of \emph{SQUAREM} through a wide application of EM/MM problems, including interval censoring, genetics admixture, and logistic regression maximum likelihood estimation (an MM problem).

\clearpage
\section{Examples}\label{examples}

\subsection{Factor analysis}\label{factor-analysis}
Factor analysis is a statistical modeling approach that aims to explain the
variability among observed variables in terms of a smaller set of unobserved factors. Factor analysis
is widely applied in areas where observed
variables may be conceptualized as manifesting from some unobserved latent
factors, such as psychometrics, behavioral sciences, social sciences,
and marketing. The latent factors can be regarded as missing
data in a multivariate normal model and the EM algorithm \citep{dempster1977maximum}, 
therefore, becomes a natural way to compute the maximum likelihood estimates.
We will illustrate using two examples the dramatic accelerations of EM by
Squarem and also compare with ECME \citep{liu1998maximum}, an acceleration of EM. 
One example comes from real data, as used by \cite{liu1998maximum} and \cite{rubin1982algorithms}, while the other is a simulation
example.

\paragraph{Notations}\label{notations}

Following the notation in \cite{rubin1982algorithms}, let \(Y\) be
\(n \times p\) observed data matrix and Z be \(n \times q\) unobserved
factor matrix where \(q \leq p\). \((Y_i, Z_i), i = 1, 2, \hdots, n\)
are independently and identically distributed vectors following multivariate normal distribution. The marginal distribution
of \(Z_i, i = 1, 2, \hdots, n\) is such that \(Z_{i_{q \times 1}} \sim\)
multivariate
normal\(\left( {\begin{pmatrix} 0 \\ 0 \\ \vdots \\ 0 \end{pmatrix}}_{q \times 1}, R_{q \times q} \right)\).
Let the variance of each component of \(Z_i, i = 1, 2, \hdots, n\) be
\(1\), so \(R\) is also the correlation matrix for \(Z_i\). Factor
analysis model assumes that given the factors \(Z_i\), the components of
vector \(Y_i\) become independent and
\(Y_{i_{p \times 1}} | Z_{i_{q \times 1}} \sim\) multivariate
normal\(\left( \alpha_{p \times 1} + \beta_{q \times p}^\top Z_{i_{q \times 1}} , \tau^2_{p \times p} \right)\)
where \(\tau^2 = \text{diag}\{\tau_1^2, \tau_2^2, \hdots, \tau_p^2\}\).
\(\beta_{q \times p}\) is called factor loading matrix while the
diagonal variances in \(\tau^2\) are called uniquenesses in factor
analysis. In general, maximum likelihood factor analysis involves
estimating \(\alpha, \beta, \tau^2\) and \(R\). But \(R\) is often
considered identity matrix (orthogonal factor model) and the maximum
likelihood estimator of \(\alpha\) is always \(\bar{Y}\), the column
means of observed data matrix \(Y\). Suppose we center matrix \(Y\) by
its column means, \(\alpha\) is always vector zero. Therefore, we are
left with only \(\beta, \tau^2\) to estimate. Given \(\beta, \tau^2\),
the marginal distribution of \(Y_i, i = 1, 2, \hdots, n\) is
multivariate normal with mean zero vector and covariance matrix
\(\tau^2 + \beta^\top \beta\). Thus we can write the log likelihood of
observed data matrix \(Y\),

\[ ll(\tau^2, \beta) = -\frac{n}{2}\log|\tau^2 + \beta^\top \beta| - \frac{n}{2} \text{tr}[C_{yy}(\tau^2 + \beta^\top \beta)^{-1}],\]
where \(C_{yy}\) is the sample covariance of \(Y\). The negative log
likelihood to be minimized is coded in function \verb+factor.loglik()+, to
check monotonicity in Squarem algorithm.

\paragraph{EM algorithm}\label{em-algorithm}
For derivation of EM step, see Appendix~\ref{facem}.

If the loading matrix $\beta$ is unrestriced, the EM update is such that 
\begin{align*}
\beta^{(k+1)} & = (\delta^\top C_{yy}\delta + \Delta)^{-1}(C_{yy}\delta)^\top ,\\
{\tau^2}^{(k+1)} & = \text{diag}\{ C_{yy} - C_{yy}\delta(\delta^\top C_{yy}\delta + \Delta)^{-1}(C_{yy}\delta)^\top \},\\
\end{align*} where $\beta^{(k+1)}, {\tau^2}^{(k+1)}$ are the $(k+1)^{th}$ iteration of estimates and $\delta, \Delta$ are defined in the Appendix~\ref{facem}.

Similarly, if the loading matrix $\beta$ has a priori zeroes, the EM update is such that 
\begin{align*}
\beta^{(k+1)}_{1j} & = {(\delta^\top C_{yy}\delta + \Delta)_{1j}}^{-1}(C_{yy}\delta)_{1j}^\top , \\
{\tau^2}^{(k+1)}_{j} & = C_{yyj} - (C_{yy}\delta)_{1j}{(\delta^\top C_{yy}\delta + \Delta)_{1j}}^{-1}(C_{yy}\delta)_{1j}^\top ,
\end{align*} 
where $j$ refers to the $j^{th}$ variable in vector $Y_i, i = 1, 2, \hdots, n$, subscript $1j$ corresponds to the factors with nonzero loadings for the $j^{th}$ variable and $C_{yyj}$ is the $j^{th}$ diagonal element of $C_{yy}$. Next we consider two data examples to illustrate the simplicity and stability of Squarem to accelerate EM algorithm.

\paragraph{Real data example}\label{real-data-example}

We use the real data from \cite{joreskog1967general} as in \cite{rubin1982algorithms} and \cite{liu1998maximum}. The data consists of 9 variables, 4 factors, and 2 patterns of a priori zeroes for the loadings such that one a priori zero loadings on factor 4 for variables 1 through 4,  and a different a priori zero loadings on factor 3 for variables 5-9. There is otherwise no restrictions. The sample covariance matrix $C_{yy}$ is given below:

$$C_{yy} = \begin{pmatrix}
1.0 & 0.554 & 0.227 & 0.189 & 0.461 & 0.506 & 0.408 & 0.280 & 0.241 \\
    & 1.0   & 0.296 & 0.219 & 0.479 & 0.530 & 0.425 & 0.311 & 0.311 \\
    &       & 1.0   & 0.769 & 0.237 & 0.243 & 0.304 & 0.718 & 0.730 \\
    &       &       & 1.0   & 0.212 & 0.226 & 0.291 & 0.681 & 0.661 \\
    &       &       &       & 1.0   & 0.520 & 0.514 & 0.313 & 0.245 \\
    &       &       &       &       & 1.0   & 0.473 & 0.348 & 0.290 \\
    &       &       &       &       &       & 1.0   & 0.374 & 0.306 \\
    &       &       &       &       &       &       & 1.0   & 0.672 \\
    &       &       &       &       &       &       &       & 1.0
\end{pmatrix}.$$

We use the starting values of $\beta$ and $\tau^2$ as in \cite{liu1998maximum}, where 

$${\beta ^{\text{start}}}^\top = \begin{pmatrix}
0.5954912 & -0.4893347 & -0.3848925  & 0.0000000 \\
0.6449102 & -0.4408213 & -0.3555598  & 0.0000000 \\
0.7630006 & 0.5053083 & -0.0535340  & 0.0000000 \\
0.7163828 & 0.5258722 & 0.0219100  & 0.0000000 \\
0.6175647 & -0.4714808 & 0.0000000  & 0.1931459 \\
0.6464100 & -0.4628659 & 0.0000000  & 0.4606456 \\
0.6452737 & -0.3260013 & 0.0000000 & -0.3622682 \\
0.7868222 & 0.3690580 & 0.0000000  & 0.0630371 \\
0.7482302 & 0.4326963 & 0.0000000  & 0.0431256
\end{pmatrix},$$ and ${\tau^2_{j}}^{\text{start}} = 10^{-8}$ for $j=1, 2, \hdots, 9.$

The negative log likelihood is given by the function \verb+factor.loglik()+ below, which corresponds to the argument \verb+objfn} in \verb+squarem()+.

\begin{verbatim}
R> factor.loglik <- function(param, cyy) {
+    beta.vec <- param[1:36]
+    beta.mat <- matrix(beta.vec, 4, 9)
+    tau2 <- param[37:45]
+    tau2.mat <- diag(tau2)
+    Sig <- tau2.mat + t(beta.mat) %*% beta.mat
+    loglik <- -145/2 * log(det(Sig)) - 145/2 * sum(diag(solve(Sig, cyy)))
+    return(-loglik)
+  }
\end{verbatim}

One EM update is given by the function \verb+factor.em()+ below, which corresponds to the argument \verb+fixptfn+ in \verb+squarem()+.

\begin{verbatim}
R> factor.em <- function(param, cyy) {
+    param.new <- rep(NA, 45)
+    beta.vec <- param[1:36]
+    beta.mat <- matrix(beta.vec, 4, 9)
+    tau2 <- param[37:45]
+    tau2.mat <- diag(tau2)
+    inv.quantity <- solve(tau2.mat + t(beta.mat) %*% beta.mat)
+    small.delta <- inv.quantity %*% t(beta.mat)
+    big.delta <- diag(4) - beta.mat %*% inv.quantity %*% t(beta.mat)
+    cyy.inverse <- t(small.delta) %*% cyy %*% small.delta + big.delta
+    cyy.mat <- t(small.delta) %*% cyy
+    beta.new <- matrix(0, 4, 9)
+    beta.p1 <- solve(cyy.inverse[1:3, 1:3]) %*% cyy.mat[1:3, 1:4]
+    beta.p2 <- solve(cyy.inverse[c(1, 2, 4), c(1, 2, 4)]) 
+      %*% cyy.mat[c(1, 2, 4), 5:9]
+    beta.new[1:3, 1:4] <- beta.p1
+    beta.new[c(1, 2, 4), 5:9] <- beta.p2
+    tau.p1 <- diag(cyy)[1:4] - diag(t(cyy.mat[1:3, 1:4]) 
+      %*% solve(cyy.inverse[1:3, 1:3]) 
+      %*% cyy.mat[1:3, 1:4])
+    tau.p2 <- diag(cyy)[5:9] - diag(t(cyy.mat[c(1, 2, 4), 5:9]) 
+      %*% solve(cyy.inverse[c(1, 2, 4), c(1, 2, 4)]) 
+      %*% cyy.mat[c(1, 2, 4), 5:9])
+    tau.new <- c(tau.p1, tau.p2)
+    param.new <- c(as.numeric(beta.new), tau.new)
+    param <- param.new
+    return(param.new)
+  }
\end{verbatim}

In order to compare with ECME algorithm as implemented by \cite{liu1998maximum}, we also write the function \verb+factor.ecme()+ below to do one ECME iteration. The only difference from EM algorithm is that for M-step, after we update the loading matrix $\beta$, we find $\tau^2$ that maximizes the actual constrained likelihood of observed data matrix $Y$ given the updated $\beta$ using Newton-Raphson.

\begin{verbatim}
R> factor.ecme <- function(param, cyy) {
+    n <- 145
+    param.new <- rep(NA, 45)
+    beta.vec <- param[1:36]
+    beta.mat <- matrix(beta.vec, 4, 9)
+    tau2 <- param[37:45]
+    tau2.mat <- diag(tau2)
+    inv.quantity <- solve(tau2.mat + t(beta.mat) %*% beta.mat)
+    small.delta <- inv.quantity %*% t(beta.mat)
+    big.delta <- diag(4) - beta.mat %*% inv.quantity %*% t(beta.mat)
+    cyy.inverse <- t(small.delta) %*% cyy %*% small.delta + big.delta
+    cyy.mat <- t(small.delta) %*% cyy
+    beta.new <- matrix(0, 4, 9)
+    beta.p1 <- solve(cyy.inverse[1:3, 1:3]) %*% cyy.mat[1:3, 1:4]
+    beta.p2 <- solve(cyy.inverse[c(1, 2, 4), c(1, 2, 4)]) 
+      %*% cyy.mat[c(1, 2, 4), 5:9]
+    beta.new[1:3, 1:4] <- beta.p1
+    beta.new[c(1, 2, 4), 5:9] <- beta.p2
+    A <- solve(tau2.mat + t(beta.new) %*% beta.new)
+    sum.B <- A %*% (n * cyy) %*% A
+    gradient <- - tau2/2 * (diag(n*A) - diag(sum.B))
+    hessian <- (0.5 * (tau2 %*% t(tau2))) * (A * (n * A - 2 * sum.B))
+    diag(hessian) <- diag(hessian) + gradient
+    U <- log(tau2)
+    U <- U - solve(hessian, gradient)
+    tau.new <- exp(U)
+    param.new <- c(as.numeric(beta.new), tau.new)
+    param <- param.new
+    return(param.new)
+  }
\end{verbatim}

Next we use \textbf{R} Package \emph{SQUAREM} to compute the MLE by EM, Squarem, ECME, and Squared ECME algorithms. Tolerance is set to be $10^{-8}$ across all algorithms.

\begin{itemize}

\item EM 

In order to perform the EM algorithm, we use function \verb+fpiter()+ in \emph{SQUAREM} Package. The arguments consist of a starting value, function \verb+factor.em()+ that encodes one EM update, negative log likelihood function \verb+factor.loglik()+, other variables as needed by these functions, and a control list to specify changes to default values. The starting value for $\beta$ and $\tau^2$ comes from \cite{liu1998maximum}.

\begin{verbatim}
R> library("SQUAREM")
R> system.time(f1 <- fpiter(par = param.start, cyy=cyy, 
+    fixptfn = factor.em, objfn = factor.loglik, 
+    control = list(tol = 10^(-8), maxiter = 20000)))
\end{verbatim}

\begin{verbatim}
 user  system elapsed
2.805   0.028   2.834
\end{verbatim}

\begin{verbatim}
R> f1$fpevals
\end{verbatim}

\begin{verbatim}
[1] 14659
\end{verbatim}

It takes 14659 iterations to converge for the EM algorithm, which spends 2.834 seconds. 

\item ECME

We replace function \verb+factor.em()+ by \verb+factor.ecme()+ that implements one ECME update thus to implement ECME algorithm.

\begin{verbatim}
R> system.time(f2 <- fpiter(par = param.start, cyy = cyy, 
+    fixptfn = factor.ecme, objfn = factor.loglik, 
+    control = list(tol = 10^(-8), maxiter = 20000)))
\end{verbatim}

\begin{verbatim}
 user  system elapsed
1.378   0.029   1.409
\end{verbatim}

\begin{verbatim}
R> f2$fpevals
\end{verbatim}

\begin{verbatim}
[1] 6408
\end{verbatim}

It takes 6408 iterations of ECME updates to converge, less than half of what the EM needs. Also it spends 1.409 seconds, approximately half of the time it takes for the EM to converge.

\item Squarem

Next, we use function \verb+squarem()+ in \emph{SQUAREM} Pakcage to apply Squarem algorithm to accelerate EM. The arguments are the same as in function \verb+fpiter()+ except a few control parameters particularly set for Squarem algorithm.

\begin{verbatim}
R> system.time(f3 <- squarem(par = param.start, cyy = cyy, 
+    fixptfn = factor.em, objfn = factor.loglik, 
+    control = list(tol = 10^(-8))))
\end{verbatim}

\begin{verbatim}
 user  system elapsed
0.226   0.006   0.233
\end{verbatim}

\begin{verbatim}
R> f3$fpevals
\end{verbatim}

\begin{verbatim}
[1] 876
\end{verbatim}

It only takes 876 iterations of EM updates to converge, which is faster by a factor of 17 and 7 in terms of the number of fixed point evaluations when compared to the EM and ECME, respectively.  Moreover, Squarem only uses 0.233 seconds to converge, 12 times faster than the EM and 6 times faster than ECME.

\item Squared ECME

Squarem algorithm can even be used to accelerate ECME, which is already a faster version of the EM algorithm. Let us call this Squared ECME.

\begin{verbatim}
R> system.time(f4 <- squarem(par = param.start, cyy = cyy, 
+    fixptfn = factor.ecme, objfn = factor.loglik, 
+    control = list(tol = 10^(-8))))
\end{verbatim}

\begin{verbatim}
 user  system elapsed
0.111   0.005   0.117
\end{verbatim}

\begin{verbatim}
R> f4$fpevals
\end{verbatim}

\begin{verbatim}
[1] 400
\end{verbatim}

The squared ECME converges in only 400 iterations compared to 6400 iterations for ECME, and it takes only 0.11 seconds.

\end{itemize}

In order to accommodate the randomness of CPU time, we run the above 4 schemes 100 times and summarize the mean and standard deviation of CPU running time into Table~\ref{faemcomp}, along with the number of fixed point evaluations needed.

\begin{table}[h!]
\centering
\begin{tabular}{lllll}
                  & EM           & ECME         & Squarem      & Squared ECME \\
                  \hline
CPU time          & 2.799(0.116) & 1.382(0.046) & 0.221(0.012) & 0.107(0.005) \\
Number of EM iterations & 14659        & 6408         & 876          & 400         
\end{tabular}
\caption{The comparison of EM, ECME, Squarem and Squared ECME algorithms on real data from \cite{joreskog1967general}. Note the CPU time is in seconds.}
\label{faemcomp}
\end{table}

Table~\ref{faemcomp} demonstrates that Squarem algorithm can greatly and easily improve on both EM and ECME algorithms for factor analysis problem.

\paragraph{Simulation example}\label{hypothetical-example}

In the simulation example, we generate 200 observations of 32 subject scores and we assume that there are 4 latent factors. In this case, we do not impose any priori zero loadings for convenience of comparison. The function used to generate data and compare the performance of EM to Squarem is coded into \verb+simulate.FAEM()+, accessible from the replication script. 

\begin{figure}[ht!]
\centering
  \subfigure[The cumulative distribution function of CPU running time for EM algorithm versus Squarem.]{
    \label{comp1}
    \begin{minipage}[b]{0.48\textwidth}
      \centering
      \includegraphics[width=\textwidth]{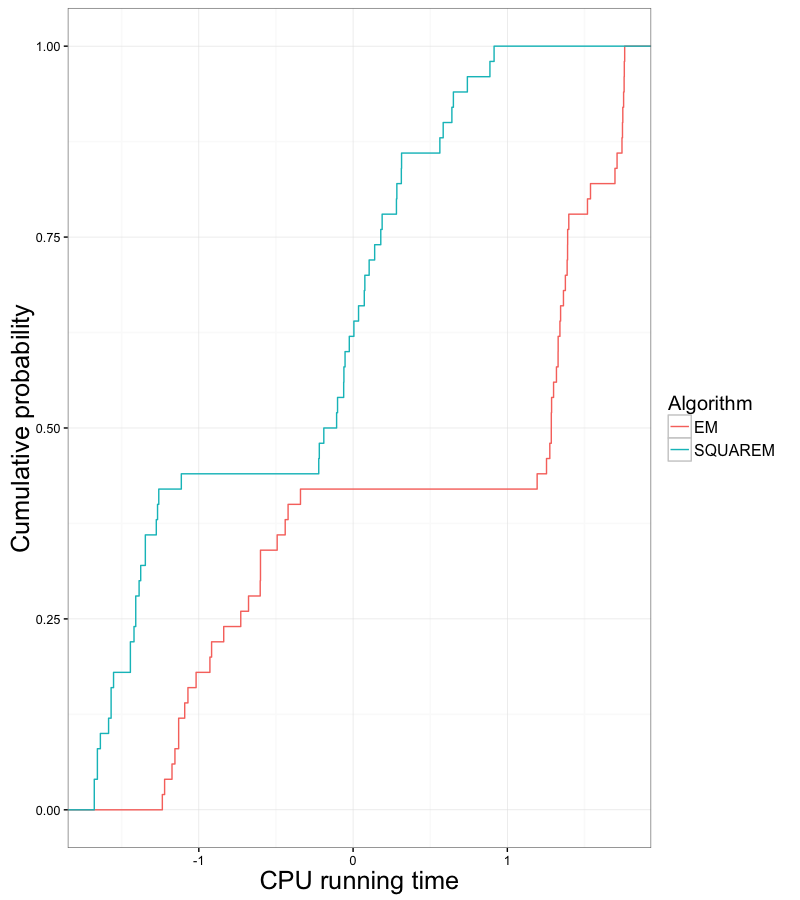}
    \end{minipage}}
  \subfigure[The cumulative distribution function of the number of EM evaluations (log10 scale) for EM algorithm versus Squarem.]{
    \label{comp2} 
    \begin{minipage}[b]{0.48\textwidth}
      \centering
      \includegraphics[width=\textwidth]{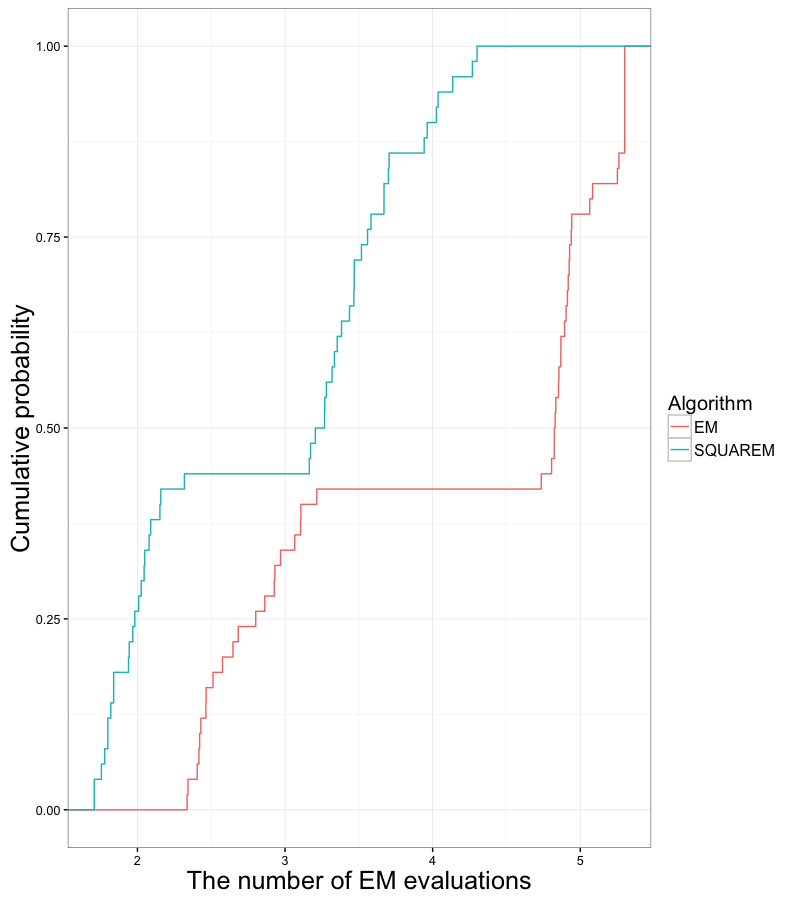}
    \end{minipage}}
  \subfigure[The scatter plot of CPU running time for EM algorithm versus Squarem.]{
    \label{comp3} 
    \begin{minipage}[b]{0.48\textwidth}
      \centering
      \includegraphics[width=\textwidth]{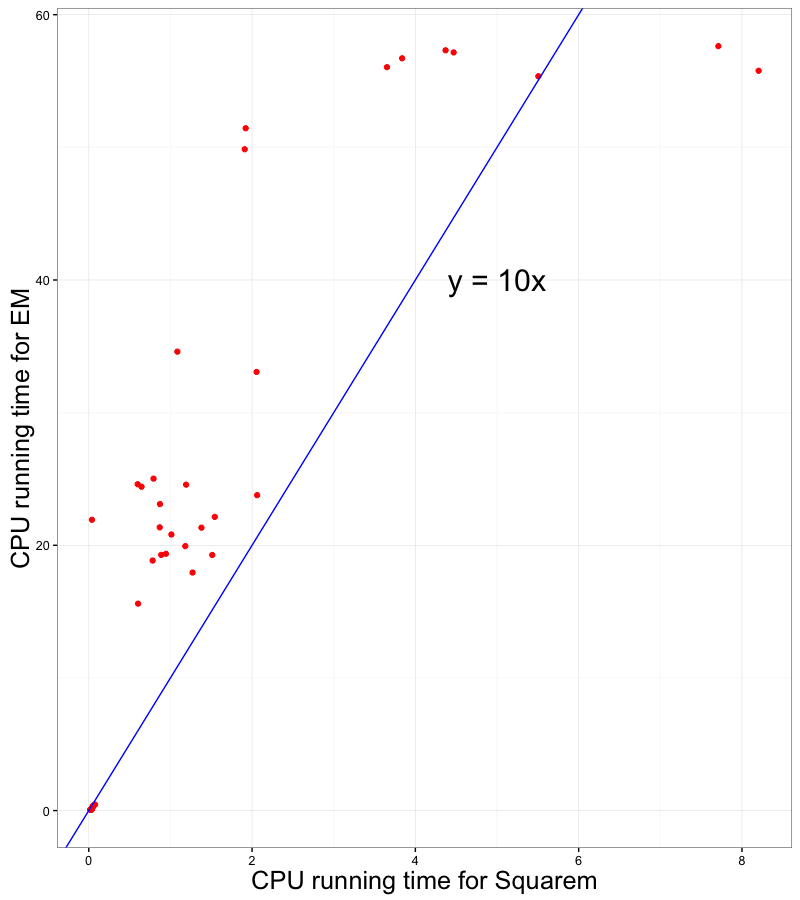}
    \end{minipage}}
  \subfigure[The scatter plot of the number of EM evaluations for EM algorithm versus Squarem.]{
    \label{comp4} 
    \begin{minipage}[b]{0.48\textwidth}
      \centering
      \includegraphics[width=\textwidth]{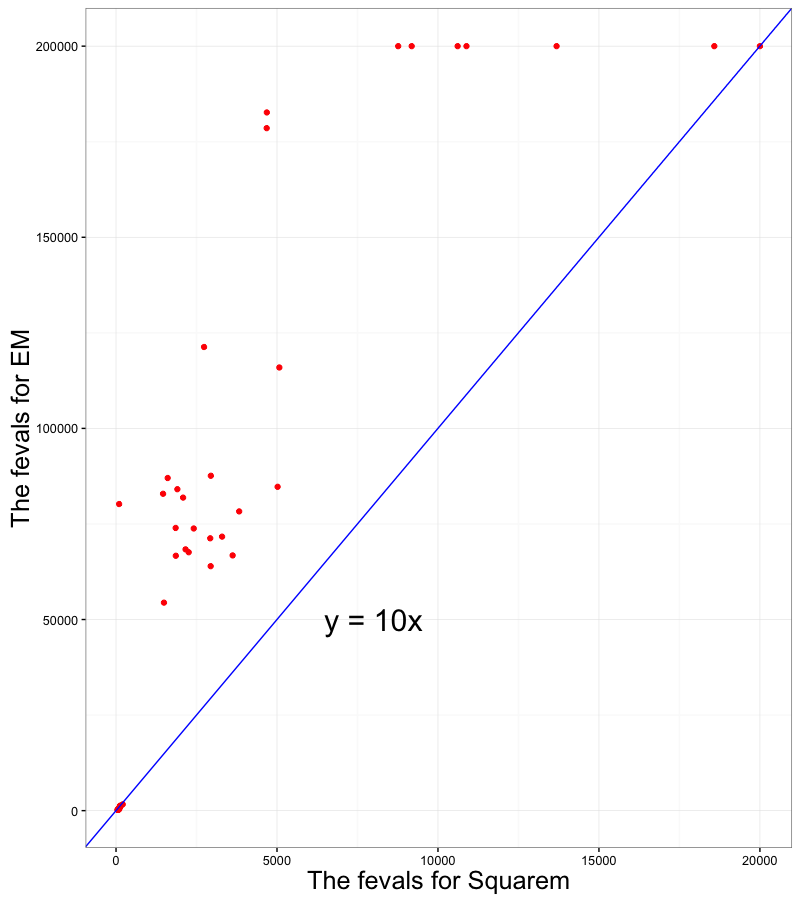}
    \end{minipage}}
\caption{The comparison of CPU running time and the number of EM evaluations between EM algorithm and Squarem.}
\label{emsqhypo}
\end{figure}

The results of comparison of CPU running time and the number of EM evaluations between EM algorithm and Squarem are summarized in Figure~\ref{emsqhypo}. It can be seen that Squarem performs consistently better than EM algorithm for both criteria, by a factor of at least 10 in most cases. 

\subsection{Interval censoring}\label{interval-censoring}
Interval censoring is a common phenomenon in survival analysis, where we do not observe the precise time of an event for each individual, but we only know the time interval during which the individual's event occurs. Following the notations in \cite{gentleman1994maximum}, we assume that survival time, $X$, also known as failure time, come from a distribution $F$. Each individual $i$ goes through a sequence of inspection times $t_{i,1},t_{i,2},\hdots$. The survival time $x_i$ for individual $i$ is not observed, however, the last inspection time prior to $x_i$ and the first inspection time after are recorded. An example of interval censored data is displayed in Table~\ref{intervalexample}.

\begin{table}[h!]  
\begin{tabular}{lcc}
                  &  Last inspection time prior to $x_i$         & First inspection time after $x_i$        \\
                  \hline
Individual 1    & 1  & 3  \\
Individual 2    & 2   & 6     \\
\vdots & \vdots & \vdots \\
Individual n    & 3   & 4
\end{tabular}
\caption{The example of interval censored data(unit: year).}
\label{intervalexample}
\end{table}

Therefore, data consists of time intervals $I_i = (L_i, R_i)$ for each individual $i , i=1,2,\hdots,n$ and the event for individual $i$ is known to happen during that interval. Let $\{s_j\}_{j=0}^{m}$ be the unique ordered times of $\{0,\{L_i\}_{i=1}^{n},\{R_i\}_{i=1}^{n}\}$, and $\alpha_{ij},i=1,2,\hdots,n,j=1,2,\hdots,m$, the $ij$ cell of an $\alpha$ matrix, be such that 

$$\alpha_{ij} = \begin{cases}
1 & \text{if $(s_{j-1}, s_{j}) \subseteq I_i$, the event for individual $i$ can occur in $(s_{j-1}, s_{j})$}\\
0 & \text{otherwise}\\
\end{cases}$$

and $p_j = F(s_j-) - F(s_{j-1})$, $p = (p_1,p_2,\hdots,p_m)^\top $. The log likelihood of data is therefore $$ll(p) = \sum_{i=1}^n \log{(\sum_{j=1}^{m} \alpha_{ij}p_j)}.$$ The negative log likelihood is coded in function \verb+loglik()+, corresponding to the argument \verb+objfn+ in \verb+squarem()+. A in function \verb+loglik()+ refers to the alpha matrix $\alpha$ and pvec is the vector of probabilities, $p$.

\begin{verbatim}
R> loglik <- function(pvec, A) {
+    -sum(log(c(A %*% pvec)))
+  }
\end{verbatim}

\paragraph{EM algorithm}\label{em-algorithmint}
For derivation of EM step, see Appendix~\ref{intsem}.

The EM update is such that 
\begin{align*}
p_j^{(k+1)} & = \frac{1}{n}\sum_i \mu_{ij}, j=1,2,\hdots,m \\
p^{(k+1)} & = (p_1^{(k+1)}, p_2^{(k+1)}, \hdots, p_m^{(k+1)})^\top
\end{align*} where $p^{(k+1)}$ is the $(k+1)^{th}$ iteration of estimates and $\mu_{ij} = \frac{\alpha_{ij}p_j}{\sum_{s} \alpha_{is}p_s}, i=1,2,\hdots,n,j=1,2,\hdots,m.$ Such one EM update is written in function \verb+intEM()+, corresponding to the argument \verb+fixptfn+ in \verb+squarem()+.

\begin{verbatim}
R> intEM <- function(pvec, A) {
+    tA <- t(A)
+    Ap <- pvec * tA 
+    pnew <- colMeans(t(Ap)/colSums(Ap))
+    pnew * (pnew > 0)
+  }
\end{verbatim}

\paragraph{EM-ICM algorithm}\label{em-icm-algorithm}

\cite{wellner1997hybrid} developed a hybrid algorithm called EM-ICM for the MLE computation of interval censored data.  This algorithm alternates steps of iterative convex minorant(ICM) and of EM. \cite{wellner1997hybrid} showed in paper that EM-ICM substantially improves the performance of the EM algorithm. We also compare Squarem with EM-ICM using real data example and a simulated one. We use function \verb+EMICM()+ in \textbf{R} Package \emph{interval}, written by \cite{michaelfay}, to implement EM-ICM algorithm.

\paragraph{Real data example}\label{real-data-exampleint}
The real data comes from \cite{finkelstein}, which gives the interval when cosmetic deterioration occurred in 46 individuals with early breast cancer under radiotherapy. Table~\ref{intrexample} shows censored interval for each individual.

\begin{table}[h!]
\centering
\begin{tabular}{cccccc}
&&&&&\\
(45,Inf] & (6,10] & (0,7] & (46,Inf] & (7,16] & (17,Inf]  \\
(7,14] & (37,44] & (0,8] & (4,11]  & (15,Inf] & (11,15] \\    
(22,Inf] & (46,Inf] & (46,Inf] & (25,37] & (46, Inf] & (26, 40]\\
(46, Inf] & (27,34] & (36,44] & (46, Inf] & (36, 48] & (37, Inf]\\
(40, Inf] & (17,25] & (46, Inf] & (11,18] & (38, Inf] & (5, 12]\\
(37, Inf] & (0,5] & (18, Inf] & (24, Inf] & (36, Inf] & (5, 11]\\
(19, 35] & (17,25] & (24, Inf] & (32, Inf] & (33, Inf] & (19,26]\\
(37,Inf] & (34,Inf] & (36, Inf] & (46,Inf] & &
\end{tabular}
\caption{The censored intervals when cosmetic deterioration occurred.}
\label{intrexample}
\end{table}

We use function \verb+Aintmap+ in R Package \emph{interval} to produce matrix $\alpha$ and then generate starting values.

\begin{verbatim}
R> library("interval")
R> A <- Aintmap(dat[, 1], dat[, 2])
R> m <- ncol(A)
R> pvec <- rep(1/m, length = m)
\end{verbatim}

We modified the function \verb+EMICM()+ in R Package \emph{interval} in order to keep the same starting values across all algorithms, a uniform starting value where $p_i = 1/m$, $i=1,2,\hdots,m$.

Next, we compare the performance of the above three algorithms, EM, Squarem and EM-ICM. We did not include EM-ICM for comparison in the number of EM evaluations because intrinsically EM-ICM algorithm is a hybrid algorithm where each EM-ICM step is different from an EM evaluation. The tolerance for convergence is set at $10^{-8}$, the same across all algorithms.

\begin{itemize}
\item EM algorithm

\begin{verbatim}
R> system.time(ans1 <- fpiter(par = pvec, fixptfn = intEM, 
+    objfn = loglik, A = A, control = list(tol = 1e-8)))
\end{verbatim}

\begin{verbatim}
 user  system elapsed
0.008   0.000   0.009
\end{verbatim}

\begin{verbatim}
ans1$fpevals
\end{verbatim}

\begin{verbatim}
[1] 216
\end{verbatim}

\item Squarem

\begin{verbatim}
R> system.time(ans2 <- squarem(par = pvec, fixptfn = intEM, 
+    objfn = loglik, A = A, control = list(tol = 1e-8)))
\end{verbatim}

\begin{verbatim}
 user  system elapsed
0.002   0.000   0.002
\end{verbatim}

\begin{verbatim}
ans2$fpevals
\end{verbatim}

\begin{verbatim}
[1] 40
\end{verbatim}

\item EM-ICM algorithm

\begin{verbatim}
R> system.time(ans3 <- EMICM.mod(dat, EMstep = TRUE, 
+    ICMstep = TRUE, keepiter = FALSE, 
+    tol = 1e-08, maxiter = 1000))
\end{verbatim}

\begin{verbatim}
 user  system elapsed
0.025   0.001   0.027
\end{verbatim}

\end{itemize}

\begin{verbatim}
R> max(abs(ans1$par - ans2$par))
\end{verbatim}

\begin{verbatim}
[1] 0
\end{verbatim}

\begin{verbatim}
R> max(abs(ans2$par - ans3$pf))
\end{verbatim}

\begin{verbatim}
[1] 4.707805e-05
\end{verbatim}

All three algorithms converge to the same point as evidenced by the maximum difference in absolute value between algorithms-produced parameter estimates. The EM algorithm performs fairly well on this real dataset, perhaps due to the small sample size.  Even so, Squarem still outperforms the EM by a factor of 5 in terms of the number of EM evaluations and a factor of 4 in CPU running time. We show in the following section that Squarem and EM-ICM algorithms are more advantageous than the EM algorithm using a simulated example as sample size increases. 
%

\paragraph{Simulation example}\label{hypothetical-exampleint}
For each individual, we randomly generate censored intervals by creating a survival time (event) and a stochastic sequence of inspection times. The left end of interval is the last inspection time before the event while the right end is the first inspection time after. The function we use to generate interval censored data is coded in \verb+gendata()+.

\begin{verbatim}
R> gendata <- function(n, mu.nexam = 5) {
+    foo <- matrix(NA, nrow = n, ncol = 3)
+    for(i in 1:n) {
+      st <- rweibull(1, shape = 1, scale = 5)
+      nexam <- rpois(1, mu.nexam)
+      exam <- round(runif(nexam, 0, 10), 1)
+      exam <- c(0, exam, Inf)
+      foo[i, ] <- c(time = st, L = max(exam[st > exam]), 
+        R = min(exam[st <= exam]))}
+    return(foo)
}
\end{verbatim}

\begin{figure}[h!]
\centering
\includegraphics[width=0.62 \textwidth]{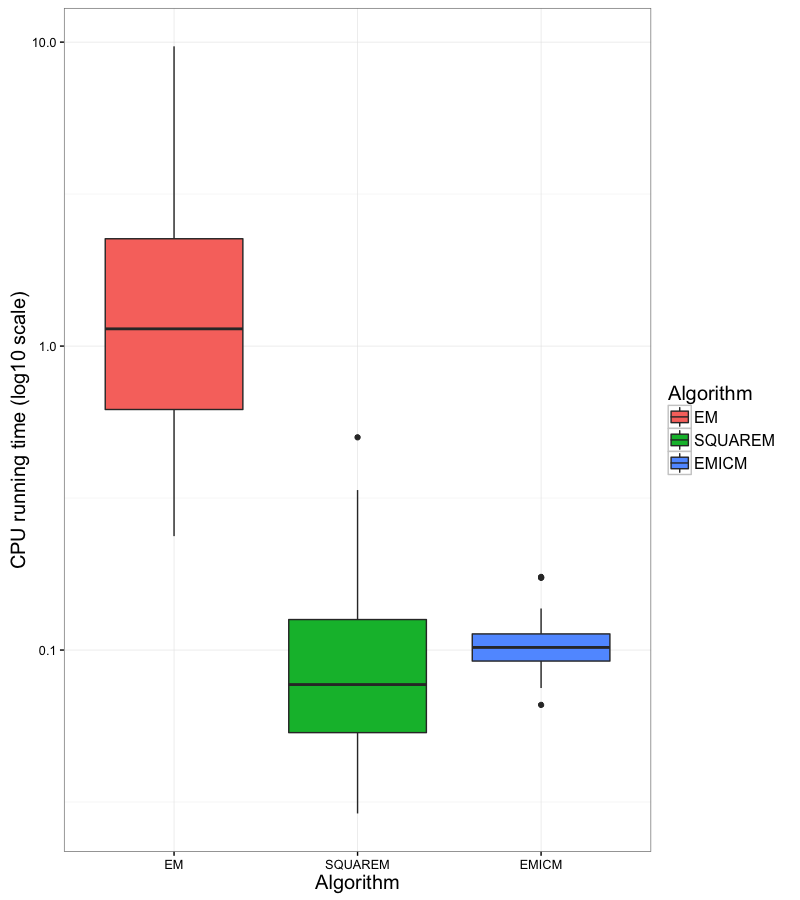}
\caption{The comparison of CPU running time among EM, Squarem and EM-ICM algorithms, applied to 100 simulated datasets for a moderate sample size $n=200$.}
\label{fig:samplesize200}
\end{figure}

First, let us try sample size $n=200$, simulate 100 times and compare the performance of the EM, Squarem and EM-ICM algorithms. The results are summarized in Figure~\ref{fig:samplesize200}.

It can be seen from Figure~\ref{fig:samplesize200} that Squarem and EM-ICM algorithms are both, on average, approximately 10 times faster than the EM, for a moderate sample size $n=200$. The performance of both algorithms are comparable. Although Squarem has a lower median CPU running time, its distribution is more variable than EM-ICM algorithm. In order to show the improvement of the number of EM evaluations for Squarem over EM algorithm, we summarize the mean and standard deviation of the number of EM evaluations for both algorithms in Table~\ref{statint} for this simulation study.

\begin{table}[h!]
\centering
\begin{tabular}{lcccc}
                  & EM           & Squarem     \\
                  \hline
Mean    & 6745      & 446         \\
Standard deviation     & 4737       & 306        
\end{tabular}
\caption{The comparison of mean and standard deviation of the number of EM evaluations between EM algorithm and Squarem on simulated data example for a moderate sample size $n=200$.}
\label{statint}
\end{table}

On average, EM algorithm takes 15 times more EM steps to converge than Squarem for this simulation study with a moderate sample size $n=200$. Next, we increase the sample size from $n=200$ to $n=2000$ and evaluate the performance of the EM, Squarem and EM-ICM algorithms again on 100 simulated interval censored datasets. 

\begin{figure}[h!]
\centering
\includegraphics[width=0.62 \textwidth]{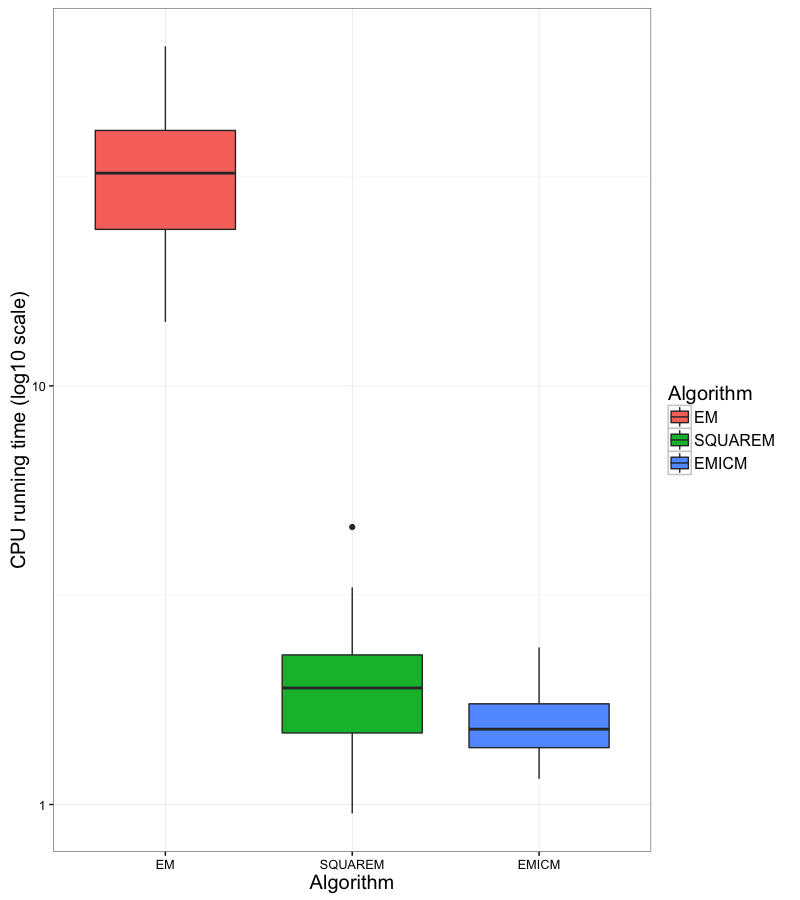}
\caption{The comparison of CPU running time among EM, Squarem and EM-ICM algorithms, applied to 100 simulated datasets for a large sample size $n=2000$.}
\label{fig:sample2000int}
\end{figure}

As sample size expands to $n=2000$, Figure~\ref{fig:sample2000int} shows that the advantage of Squarem and EM-ICM algorithms becomes greater. Both algorithms on average converge at least 17 times faster than the EM algorithm. EM-ICM is specifically tailored to interval censoring maximum likelihood estimation, hence it is not surprising that it outperforms the EM algorithm by a large margin. However, it is noteworthy that Squarem, which is a general purpose EM-like algorithm accelerator, is very competitive with EM-ICM algorithm as shown in this simulation study. Table~\ref{statint2} again compares the number of EM evaluations between EM algorithm and Squarem. Squarem on average outperforms EM algorithm by a factor of 17 in terms of the number of EM evaluations. 

\begin{table}[h!]
\centering
\begin{tabular}{lcccc}
                  & EM           & Squarem        \\
                  \hline
Mean     & 13397       & 792           \\
Standard deviation     & 3947       & 262          
\end{tabular}
\caption{The comparison of mean and standard deviation of the number of EM evaluations between EM algorithm and Squarem on simulated data example for a large sample size $n=2000$.}
\label{statint2}
\end{table}

\subsection{Genetics global ancestry estimation problem}\label{genetics-problem}
Here we demonstrate the use of Squarem to solve an important problem in quantitative genetics that is notoriously computationally challenging. Suppose our study population is an admixed population with $K$ ancestral populations. The goal is to estimate the proportion of ancestry from each contributing population for each individual's entire genome and simultaneously estimate the allele frequencies of the $K$ ancestral populations. Let us use $q_i = (q_{i1}, q_{i2}, \hdots, q_{iK})^\top$ to denote such admixture proportions for individual $i$, $i=1,2,\hdots,n$ where $q_{ik}$ is the proportion of subject $i$ genome that is attributed to the ancestral population $k$,$k=1,2,\hdots,K$ and $n$ is the number of subjects. Let Q be the $n \times K$ admixture proportions matrix. We assume that all $p$ genome-wide markers are bi-allelic (either allele 1 or allele 2). Let F be the $p \times K$ population allele frequency matrix with $f_{jk}$ being the frequency of allele 1 at marker $j$, $j=1,2,\hdots,p$ in population $k$. Matrix F and Q consist of parameters we are interested in estimating. The data consists of genetic polymorphism data sampled from $n$ diploid individuals. Specifically, we have recorded the genotype at $p$ genetic polymorphisms ("markers") for each individual. Genotype at marker $j$ for individual $i$ is represented as allele 1 counts, $x_{ij}=0,1,2$. We assume that individuals are independent and under ADMIXTURE model, the log likelihood of data is:
$$ll(F,Q) = \sum_i^{n} \sum_j^{p} (x_{ij}\log{\sum_k^{K} q_{ik}f_{jk}} + (2 - x_{ij})\log{\sum_k^{K} q_{ik}(1-f_{jk}})) + C$$ where $C$ is a constant that does not contain parameters from F and Q. See \cite{alexander2009fast} for full description of model.

The negative log likelihood of data is coded in function \verb+loglike()+, corresponding to the argument \verb+objfn+ in \verb+squarem()+.

\begin{verbatim}
R> loglike <- function(param, X, K) {
+    n <- nrow(X)
+    p <- ncol(X)
+    F <- matrix(param[1: (p * K)], p, K)
+    Q <- matrix(param[(p * K + 1): (p * K + n * K)], n, K)
+    loglikelihood <- sum(X * log(Q %*% t(F)) + (2 - X) * 
+      log(Q %*% (1 - t(F))))
+    return(-loglikelihood)
+  }
\end{verbatim}

\paragraph{EM algorithm}\label{em-adm}
For derivation of EM step, see Appendix~\ref{genem}.

The EM update of matrix F and Q is such that 
\begin{align*}
f_{jk} &= \frac{n_{jk}^{(1)}}{n_{jk}^{(1)} + n_{jk}^{(0)}},\\ 
q_{ik} &= \frac{m_{ik}}{\sum_k m_{ik}},
\end{align*} where $n_{jk}^{(1)}, n_{jk}^{(0)}, m_{ik}$ are defined in the Appendix~\ref{genem}.

This one EM evaluation is written in function \verb+admixture.em()+, corresponding to the argument \verb+fixptfn+ in \verb+squarem()+. We adapt the code provided on Peter Carbonetto github account \citep{petergit}.

\begin{verbatim}
R> admixture.em <- function(param, X, K) {
+    eps <- 1e-6
+    n <- nrow(X)
+    p <- ncol(X)
+    m  <- matrix(eps, n, K)
+    n0 <- matrix(eps, p, K)
+    n1 <- matrix(eps, p, K)
+    F <- matrix(param[1: (p * K)], p, K)
+    Q <- matrix(param[(p * K + 1): (p * K + n * K)], n, K)
+    r <- array(0, dim = c(p, 4, K, K))
+    for (i in 1:n) {
+      colnames(r) <- c("00","01","10","11")
+        for (j in 1:K) 
+          for (k in 1:K) {
+            r[,"00", j, k] <- (X[i, ] == 0) * 
+              (1 - F[, j]) * (1 - F[, k])
+            r[,"01", j, k] <- (X[i, ] == 1) * 
+              (1 - F[, j]) * F[, k]
+            r[,"10", j, k] <- (X[i, ] == 1) * 
+              F[, j] * (1 - F[, k])
+            r[,"11", j, k] <- (X[i, ] == 2) * 
+              F[, j] * F[, k]
+            r[, , j, k]     <- r[, ,j,k] * 
+              Q[i, j] * Q[i, k]
+          }
+      dim(r) <- c(p, 4 * K^2)
+      r <- r / rowSums(r)
+      dim(r) <- c(p, 4, K, K)
+      colnames(r) <- c("00", "01", "10", "11")
+      m[i, ] <- m[i, ] + apply(r, 3, sum) + apply(r, 4, sum) 
+      for (k in 1:K) { 
+        n0[, k] <- n0[, k] + rowSums(drop(r[, "00", k, ])) + 
+          rowSums(drop(r[, "01", k, ])) + 
+          rowSums(drop(r[, "00", , k])) + 
+          rowSums(drop(r[, "10", , k]))
+        n1[, k] <- n1[, k] + rowSums(drop(r[, "10", k, ])) + 
+          rowSums(drop(r[, "11", k, ])) + 
+          rowSums(drop(r[, "01", , k])) + 
+          rowSums(drop(r[, "11", , k]))
+      }
+    }
+    F <- n1/(n0 + n1)
+    Q <- m/rowSums(m)
+    return(c(as.vector(F), as.vector(Q)))
+  } 
\end{verbatim}

\paragraph{Simulation example}\label{hypoegadmix}
We simulate an allele 1 count matrix $X$ where there are 150 individuals and 100 markers for each individual. We use $K=3$. The starting value of $f_{jk}$ is randomly drawn from a uniform distribution in the range of $(0,1)$ while that of $q_{ik}$ is $\frac{1}{K}$. We implement the EM algorithm and Squarem to compute maximum likelihood estimates of matrix F and Q and compare their performance.

\begin{itemize}
\item EM algorithm

\begin{verbatim}
R> load("geno.sim.RData")
R> set.seed(413)
R> p <- 100
R> n <- 150
R> K <- 3
R> F <- matrix(runif(p * K), p, K)
R> Q <- matrix(1/K, n, K)
R> param.start <- c(as.vector(F), as.vector(Q))
R> system.time(f1 <- fpiter(par = param.start, fixptfn = admixture.em, 
+    objfn = loglike, control = list(tol = 1e-4), X = geno, K = 3))
\end{verbatim}

\begin{verbatim}
  user   system elapsed
197.094   5.817 203.040
\end{verbatim}

\begin{verbatim}
R> f1$fpevals
\end{verbatim}

\begin{verbatim}
[1] 1115
\end{verbatim}

\item Squarem

\begin{verbatim}
R> system.time(f2 <- squarem(par = param.start, fixptfn = admixture.em, 
+    objfn = loglike, 
+    control = list(tol = 1e-4, maxiter = 2000), X = geno, K = 3))
\end{verbatim}

\begin{verbatim}
 user   system  elapsed
47.460   1.403  48.869
\end{verbatim}

\begin{verbatim}
R> f2$fpevals
\end{verbatim}

\begin{verbatim}
[1] 270
\end{verbatim}

\end{itemize}

In this example, Squarem outperforms the EM algorithm by a factor of 4 in terms of CPU running time as well as the number of EM evaluations. For large genetic datasets, the E-step is by far the most computationally intensive part of the algorithm. For a faster implementation of the E-step using C (and interfaced to R using the \verb+.Call()+ function), see \cite{petergit}. Although the admixutre problem is naturally framed using EM, its convergence is very slow. \cite{alexander2009fast} implemented faster solution to this problem (using a block relaxation optimization method), which has permitted application of ADMIXTURE to very large genetic datasets. Our EM-based implementation in R is much slower than the ADMIXTURE software, but, nevertheless it serves to illustrate the benefits of Squarem in a difficult optimization problem from genetics.

\subsection{MM algorithm - logistic regression maximum likelihood estimation}\label{logistics-estimation}
In this section, we discuss a quadratic majorization algorithm (an MM algorithm) for computing the maximum likelihood estimates of logistic regression coefficients.  Minorize and maximize or equivalently, majorize and minimize (MM) algorithms typically exhibit slow linear convergence just like the EM algorithm. We show that Squarem can provide significant acceleration of MM algorithms. 

EM algorithm may be viewed as a special case of MM algorithms\citep{zhou2012vs}. The majorization algorithms are widely applied, for example, in the work of \cite{de1994block}, \cite{heiser1995convergent}, \cite{lange2000optimization}, among others. Suppose we want to minimize function $f$ over $X \subseteq \mathbb{R}^{n}$. We construct a majorization function $g$ on $X \times X$ such that $$f(x) \leq g(x, x^{(k)}) \quad \forall x, x^{(k)} \in X,$$ $$f(x^{(k)}) = g(x^{(k)}, x^{(k)}) \quad \forall x^{(k)} \in X,$$ where $k$ denotes the $k^{th}$ iteration, $k=0,1,\hdots$. Therefore, instead of minimizing $f$, we minimize $g$ such that $$x^{(k+1)} = \arg\!\min_{x \in X}{g(x, x^{(k)})}.$$ We repeat the updates of $x$ until convergence and this completes the majorization algorithm. Note that in the EM algorithm, the $Q(\theta; \theta_k)$ function plays the role of the minorizing function. 

\paragraph{Quadratic majorization algorithm}\label{mmalgorithm}
Taylor's theorem often leads to quadratic majorization algorithms \citep{bohning1988monotonicity} where the majorization function $g$ is quadratic. By Taylor's theorem, expand $f(x)$ at $x^{(k)}$, $$f(x) = f(x^{(k)}) + (x- x^{(k)} )^\top \partial f( x^{(k)} ) + \frac{1}{2}(x- x^{(k)})^\top \partial^{2}f(\xi)(x- x^{(k)} )$$ where $\xi$ is on the line between $x$ and $ x^{(k)}$. The majorization function $g$ is constructed by constructing a matrix $B$ such that $B - \partial^{2}f( \xi )$ is always positive semi-definite regardless of $\xi$. So, $$g(x, x^{(k)}) = f(x^{(k)}) + (x- x^{(k)} )^\top \partial f( x^{(k)} ) + \frac{1}{2}(x- x^{(k)})^\top B(x- x^{(k)} )$$ is a majorization function for $f$. Let us define a clever variable $z^{(k)}$ such that $z^{(k)} = x^{(k)} - B^{-1}\partial f(x^{(k)})$ and majorization function $g$ is equivalent to the following:$$g(x, x^{(k)}) = f(x^{(k)}) + \frac{1}{2}(x-z^{(k)})^\top B(x-z^{(k)})-\frac{1}{2}\partial f(x^{(k)})^\top B^{-1}\partial f(x^{(k)}).$$ At the $k^{th}$ iteration, to minimize $g(x, x^{(k)})$ over $x \in X$ is simply to minimize $(x-z^{(k)})^\top B(x-z^{(k)})$, thus the majorization algorithm becomes: $$x^{(k+1)} = x^{(k)} - B^{-1}\partial f(x^{(k)}).$$ Therefore, in order to implement quadratic majorization algorithm, we need to construct the matrix $B$ and compute the gradient of function $f$. Let us consider logistic regression maximum likelihood estimation. Suppose we have an $n \times p$ design matrix $X$ where there are $n$ subjects and $p$ predictors. Let $y_i$ be the number of successes for subject $i$ ,$i=1,2,\hdots,n$ given the overall number of experiments, $N_i$. We use $\beta$ to denote the regression coefficients. The goal is to derive the maximum likelihood estimates of $\beta$. 

The negative log likelihood of data is:
\begin{align*}
f(\beta) & = -\log{(\prod_i P(y_i))}\\
& \propto -\log{(\prod_i p_i(\beta)^{y_i}(1-p_i(\beta))^{(N_i-y_i)})}\\
& = -\sum_i y_i\log{p_i(\beta)} - \sum_i(N_i - y_i)\log{(1-p_i(\beta))}\\
& = - \sum_i y_ix_i^\top \beta - \sum_i N_i\log{(1-p_i(\beta))},
\end{align*}

where $$p_i(\beta) = \frac{1}{1 + \exp{(-x_i^\top \beta)}}.$$

The gradient of $f(\beta)$ is such that:
$$\partial f(\beta) = \sum_i (N_ip_i(\beta) - y_i)x_i = X^\top u(\beta),$$

where the $i^{th}$ element of $u(\beta)$ is $N_ip_i(\beta) - y_i$, while the second derivative of $f(\beta)$ is:
$$\partial^{2} f(\beta) = \sum_i (N_ip_i(\beta)(1-p_i(\beta)))x_ix_i^\top = X^\top V(\beta)X,$$

where $V(\beta)$ is a diagonal matrix with the $i^{th}$ diagonal element $N_ip_i(\beta)(1-p_i(\beta)).$
Based on the fact that $p_i(\beta)(1-p_i(\beta)) \leq \frac{1}{4}$, the matrix $B$ can be constructed such that $B = \frac{1}{4}X^\top NX$ where $N$ is the diagonal matrix consisting of elements $N_i$. Thus the quadratic algorithm becomes:
$$\beta^{(k+1)} = \beta^{(k)} - 4(X^\top NX)^{-1}X^\top u(\beta).$$
Let us denote the above algorithm by uniform bound quadratic algorithm since $p_i(\beta)(1-p_i(\beta)) \leq \frac{1}{4}$ uniformly for any $\beta$ and each subject $i$. \cite{jaakkola2000bayesian} and \cite{groenen2003weighted} developed a non-uniform bound, $X^\top W(\beta)X$, where $W(\beta)$ is a diagonal matrix that consists of elements $w_i(\beta) = N_i \frac{2p_i(\beta) - 1}{2 x_i^\top \beta}$, $i=1,2,\hdots,n$. Thus the non-uniform bound quadratic algorithm becomes: 
$$\beta^{(k+1)} = \beta^{(k)} - (X^\top W(\beta)X)^{-1}X^\top u(\beta).$$

\paragraph{Real data example}\label{real-data-mm}
We use the cancer remission data in \cite{lee1974computer}. The outcome is a binary indicator of whether cancer remission occurred for the subject. Column 1 is the intercept and variables $V2$, $V3$, $\hdots$, $V7$ are results of six medical tests. The first five lines of data are as follows:

\begin{verbatim}
R> ld <- read.table("lee_data.txt")
R> head(ld, 5)
\end{verbatim}

\begin{verbatim}
V1  V2   V3   V4  V5    V6    V7 V8
 1 0.8 0.83 0.66 1.9 1.100 0.996  1
 1 0.9 0.36 0.32 1.4 0.740 0.992  1
 1 0.8 0.88 0.70 0.8 0.176 0.982  0
 1 1.0 0.87 0.87 0.7 1.053 0.986  0
 1 0.9 0.75 0.68 1.3 0.519 0.980  1
\end{verbatim}

The negative log likelihood function $f(\beta)$ is coded in \verb+binom.loglike()+, corresponding to the argument \verb+objfn} in \verb+squarem()+.

\begin{verbatim}
R> binom.loglike <- function(par, Z, y) {
+    zb <- c(Z %*% par)
+    pib <- 1 / (1 + exp(- zb))
+    return(as.numeric(-t(y) %*% (Z %*% par) - sum(log(1 - pib))))
+  }
\end{verbatim}

The uniform bound quadratic majorization algorithm update and the non-uniform one are written in function \verb+qmub.update()+, \verb+qmvb.update()+, respectively, corresponding to the argument \verb+fixptfn+ in \verb+squarem()+.

\begin{verbatim}
R> qmub.update <- function(par, Z, y) {
+    Zmat <- solve(crossprod(Z)) %*% t(Z)
+    zb <- c(Z %*% par)
+    pib <- 1 / (1 + exp(-zb))
+    ub <-  pib - y
+    par <- par - 4 * c(Zmat %*% ub)
+    par
+  }
\end{verbatim}

\begin{verbatim}
R> qmvb.update <- function(par, Z, y) {
+    zb <- c(Z %*% par)
+    pib <- 1 / (1 + exp(-zb))
+    wmat <- diag((2 * pib - 1)/(2 * zb))
+    ub <-  pib - y
+    Zmat <- solve(t(Z) %*% wmat %*% Z) %*% t(Z)
+    par <- par - c(Zmat %*% ub)
+    par
+  }
\end{verbatim}

Now let us apply these two quadratic majorization algorithms and their Squared versions where we implement squared iterative scheme, to compare their performance. The tolerance used is $10^{-7}$ and the starting value is $\beta^{(0)} = (10, 10, \hdots, 10)^\top.$

\begin{itemize}
\item Uniform bound quadratic majorization algorithm

\begin{verbatim}
R> library("SQUAREM")
R> Z <- as.matrix(ld[, 1:7])
R> y <- ld[, 8]p0 <- rep(10, 7)
R> system.time(ans1 <- fpiter(par = p0, fixptfn = qmub.update, 
+    objfn = binom.loglike, 
+    control = list(maxiter = 20000), Z = Z, y = y))
\end{verbatim}

\begin{verbatim}
 user  system  elapsed
0.051   0.003   0.055
\end{verbatim}

\begin{verbatim}
R> ans1
\end{verbatim}

\begin{verbatim}
$par
[1]  58.0384838  24.6615508  19.2935824 -19.6012695   3.8959635   
[6]  0.1510923  -87.4339059 

$value.objfn 
[1] 10.87533

$fpevals 
[1] 1127 

$objfevals 
[1] 0 

$convergence
[1] TRUE
\end{verbatim}

\item Squared uniform bound quadratic majorization algorithm

\begin{verbatim}
R> system.time(ans2 <- squarem(par = p0, fixptfn = qmub.update, 
+    objfn = binom.loglike, Z = Z, y = y))
\end{verbatim}

\begin{verbatim}
 user  system  elapsed
0.011   0.000   0.011
\end{verbatim}

\begin{verbatim}
R> ans2
\end{verbatim}

\begin{verbatim}
$par
[1]  58.0384863  24.6615466  19.2935777 -19.6012645   3.8959634   
[6]  0.1510923  -87.4339043

$value.objfn
[1] 10.87533

$iter
[1] 41

$fpevals
[1] 118

$objfevals
[1] 43

$convergence
[1] TRUE
\end{verbatim}

\item Non-uniform bound quadratic majorization algorithm

\begin{verbatim}
R> system.time(ans3 <- fpiter(par = p0, fixptfn = qmvb.update, 
+    objfn = binom.loglike, 
+    control = list(maxiter = 20000), Z = Z, y = y)
\end{verbatim}

\begin{verbatim}
 user  system  elapsed
0.029   0.001   0.030
\end{verbatim}

\begin{verbatim}
R> ans3
\end{verbatim}

\begin{verbatim}
$par
[1]  58.0384866  24.6615451  19.2935760 -19.6012627   3.8959634   
[6]  0.1510923  -87.4339030

$value.objfn
[1] 10.87533

$fpevals
[1] 442

$objfevals
[1] 0

$convergence
[1] TRUE
\end{verbatim}

\item Squared non-uniform bound quadratic majorization algorithm

\begin{verbatim}
R> system.time(ans4 <- squarem(par = p0, fixptfn = qmvb.update, 
+    objfn = binom.loglike, Z = Z, y = y))
\end{verbatim}

\begin{verbatim}
 user  system  elapsed
0.009   0.000   0.009
\end{verbatim}

\begin{verbatim}
R> ans4
\end{verbatim}

\begin{verbatim}
$par
[1]  58.0384868  24.6615443  19.2935751 -19.6012618   3.8959633   
[6]  0.1510923  -87.4339024

$value.objfn
[1] 10.87533

$iter
[1] 30

$fpevals
[1] 88

$objfevals
[1] 30

$convergence
[1] TRUE
\end{verbatim}

\end{itemize}

All four algorithms converge to the same maximum likelihood estimates but Squarem improves on both uniform and non-uniform bound quadratic majorization algorithms in terms of the number of quadratic majorization updates and CPU running time (in seconds). For uniform bound, Squarem converges around 5 times faster and saves the number of quadratic majorization updates by a factor of 10. The non-uniform bound quadratic majorization improves on the uniform bound one, but its Squared version provides further acceleration. Compared to non-uniform bound algorithm, Squarem shortens the computing time by a factor of 3 and cuts the number of quadratic majorization updates by a factor of 5.

\begin{figure}[h!]
\centering
\includegraphics[width=0.92 \textwidth]{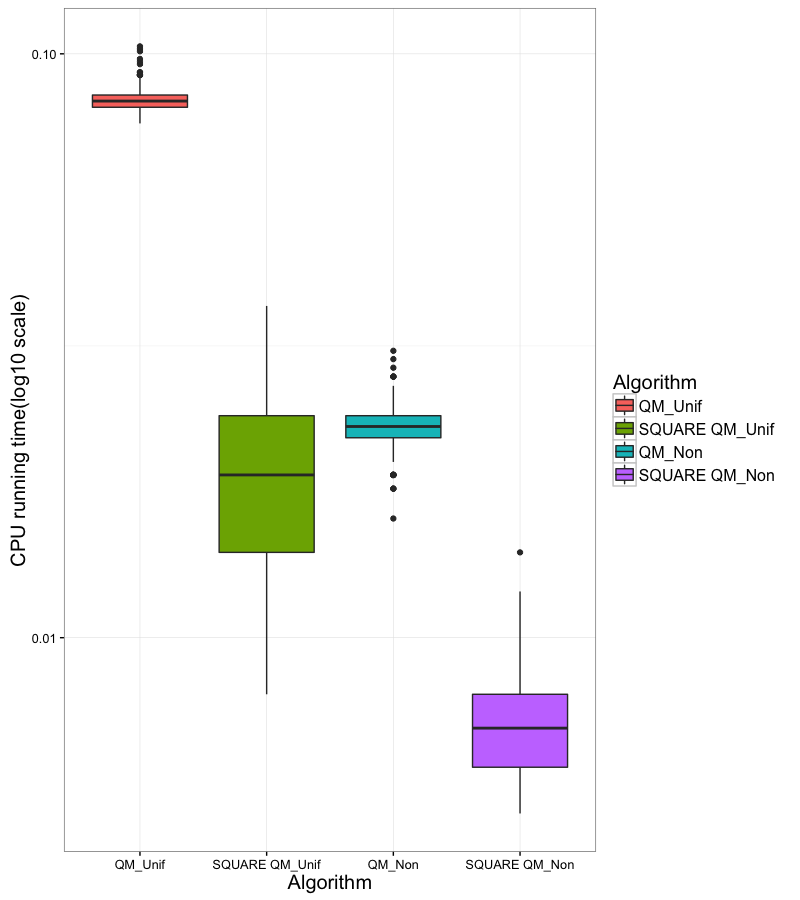}
\caption{The comparison among basic and Squared uniform bound quadratic majorization algorithms, and basic and Squared non-uniform bound quadratic majorization algorithms in terms of CPU running time (in seconds).}
\label{fig:mmres2}
\end{figure}

\begin{figure}[h!]
\centering
\includegraphics[width=0.92 \textwidth]{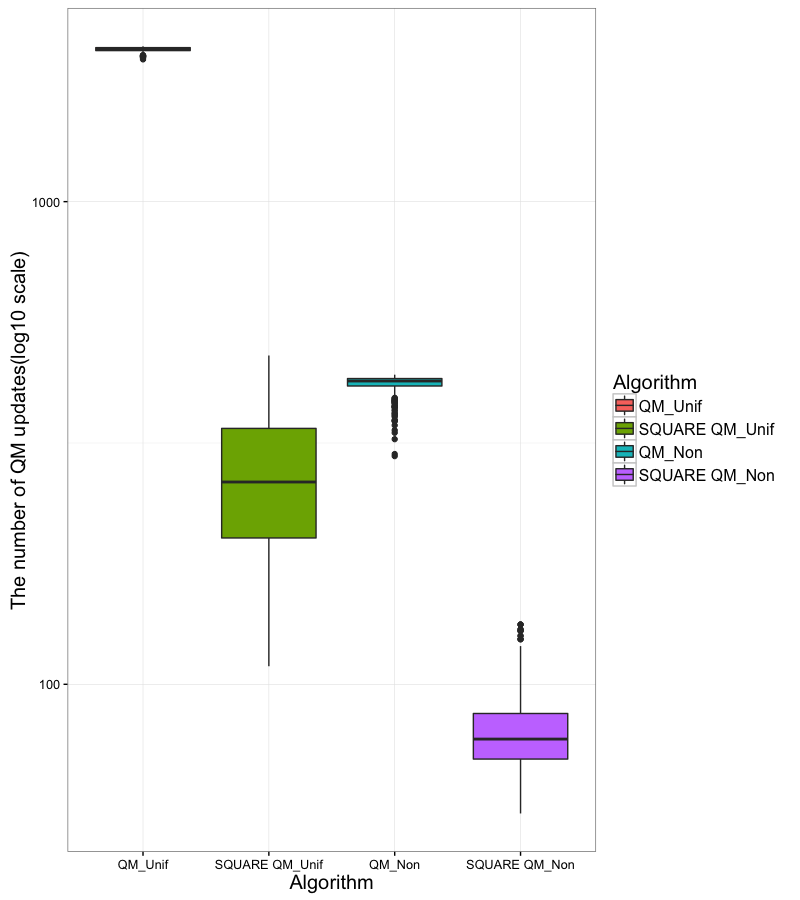}
\caption{The comparison among basic and Squared uniform bound quadratic majorization algorithms, and basic and Squared non-uniform bound quadratic majorization algorithms in terms of the number of quadratic majorization(QM) updates.}
\label{fig:mmres22}
\end{figure}

We randomly generate 500 starting values $\beta^{(0)} = (U(0,10), U(0,10), \hdots, U(0,10))^\top $ where $U(0,10)$ is a uniform random variable in the range of $(0,10)$. We then summarize the performance for these four algorithms in terms of CPU running time and the number of QM (quadratic majorization) updates in Figure~\ref{fig:mmres2} and Figure~\ref{fig:mmres22}.

Figure~\ref{fig:mmres2} and Figure~\ref{fig:mmres22} clearly show that Squarem provides substantial acceleration for both uniform bound and non-uniform bound quadratic majorization algorithms. Table~\ref{statmm} displays that for different starting values, on average, compared to the uniform bound QM algorithm, its corresponding Squared version saves the number of QM updates by a factor of 7 and runs 5 times faster while compared to the non-uniform bound QM algorithm, an already faster version over the uniform bound QM algorithm, the Squared version further improves by a factor of 3 in CPU running time and a factor of 5 in the number of QM updates. 

\begin{table}[h!]
\centering
\begin{tabular}{lcccc}
                  & Ub QM           & Squared Ub QM          & Non-Ub QM      & Squared Non-Ub QM \\
                  \hline
CPU time(mean)  & 0.079       & 0.017           & 0.021       & 0.006 \\
CPU time(sd)     & 0.006       & 0.005           & 0.003       & 0.001 \\
QM updates(mean) & 2068      & 273             & 419         & 80 \\
QM updates(sd) & 22        & 81              & 19          & 15 
\end{tabular}
\caption{The comparison among basic and Squared uniform bound quadratic majorization algorithms, and basic and Squared non-uniform bound quadratic majorization algorithms in terms of the mean and standard deviation of CPU running time and the number of quadratic majorization (QM) updates for cancer remission data with 500 randomly selected starting values.}
\label{statmm}
\end{table}

\clearpage
\section{Discussion}
Since the seminal work of \cite{dempster1977maximum}, EM and its variants have become the workhorse of computational statistics. More broadly, there are iterative algorithms which do not fit into the missing-data framework, but which are EM-like in the sense that they exhibit slow, monotone, global convergence like the EM algorithm. These include the minorize and maximize (MM) algorithm.  Even more broadly, we can include all fixed-point iterations which are contractive \citep{ortega1970iterative} and linearly convergent.  They are broader in the sense that there need not be an objective function (e.g., log-likelihood) associated with the contraction mapping.  The remarkable fact is that it is extremely easy to use Squarem to try and accelerate these iterative algorithms.  All that the user has to do is to create a function, \verb+fixptfn()+, that implements a single step of the fixed-point iteration, whether it is EM/ECM/ECME/GEM/MM or any other contractive mapping.  The objective function, \verb+objfn()+, is optional, although we recommend that it be provided, if it is easy to code. The convergence criterion used in function \verb+squarem()+ is stringent for high-dimensional problems, and in future versions, we will incorporate other parameter based criteria and criteria based on the objective function. In addition, there are other features of Squarem not illustrated in this paper, including the options of higher-order Squarem schemes and the tracking of algorithm's progress. For full features of Squarem, see \url{https://cran.r-project.org/web/packages/SQUAREM/SQUAREM.pdf}.

The main theme of the paper is that existing modeling problems based on EM-like algorithms can potentially be made computationally more efficient by using the convergence acceleration provided by Squarem. This is particularly true in high-dimensional problems where EM-like algorithms can be excruciatingly slow. There are several examples in the literature where Squarem has been effectively used. To name a few, Matthew Stephens lab at University of Chicago has been using \emph{SQUAREM} in many of their models pertaining to genetic studies where a very large number of parameters are estimated \citep[among others]{shiraishi2015simple, raj2015mscentipede}. \cite{patro2014sailfish} incorporated Squarem in the development of their new computational method, "Sailfish", to substantially increase the efficiency of processing sequencing reads. More recently, \cite{chiou2017semiparametric} have used \emph{SQUAREM} to speed up the estimation in a semi-parametric model for panel recurrent event count data. There are more such examples.  

Squarem is computationally efficient. It requires little effort beyond that of the basic fixed-point iteration.  The computation of Squarem parameter update is trivially easy since it only requires a couple of vector products.  The only place where some inefficiency could occur is in the evaluation of the objective function to assess whether the Squarem step can be accepted.  When the Squarem step results in non-monotonicity, it is rejected and instead the most recent EM update is retained. This results in some wasted effort. This is typical for algorithms with fast local convergence, for example, the Newton's method in unconstrained optimization which needs to be safe-guarded with line-search or trust-region approach to ensure global convergence. However, Squarem is efficient when the objective function evaluation is relatively cheaper than that of the fixed-point iteration, which is often the case.

Squarem can be used off-the-shelf since there is no need for the user to tweak any control parameters to optimize its performance. Given its ease of application, Squarem may be considered as a default accelerator for EM-like algorithms. We invite the reader to try Squarem acceleration on his/her own EM-like algorithm or any slowly convergent, contractive fixed-point iteration. 

\clearpage
\begin{appendices}

\section{Derivation of EM}
\subsection{Poisson mixtures}\label{poisem}
Let us define the missing variable $Z_i$,$i=0,1,\hdots,9$ such that 
$$Z_{i} = \begin{cases}
1 & \text{if death number $i$ comes from population 1}\\
0 & \text{if death number $i$ comes from population 2}\\
\end{cases}.$$ 

Let $P(Z_i = 1) = p$,$i=0,1,\hdots,9$. Given $Z_i = 1$, the death number $i$ comes from population 1, following Poisson distribution with mean $\mu_1$ while otherwise, the death number $i$ comes from population 2, following Poisson distribution with mean $\mu_2$.

\begin{itemize}

\item E-step
\begin{align*}
& Q(p, \mu_1, \mu_2 | p^{(k)}, \mu_1^{(k)}, \mu_2^{(k)}) \\
& = E[\log{[\prod_i P(i, n_i, Z_i)]} | p^{(k)}, \mu_1^{(k)}, \mu_2^{(k)}, i, n_i] \\
& = E[\log{[\prod_i ((pe^{-\mu_1}\mu_1^i/i!)^{Z_i}((1-p)e^{-\mu_2}\mu_2^i/i!)^{1-Z_i})^{n_i}]} | p^{(k)}, \mu_1^{(k)}, \mu_2^{(k)}, i, n_i] \\
& = \sum_i n_i[p_i^{(k)}(\log{p} + \log{(e^{-\mu_1}\mu_1^i/i!)}) + (1-p_i^{(k)})(\log{(1-p)} + \log{(e^{-\mu_2}\mu_2^i/i!)})],
\end{align*}

where 

\begin{align*}
p_i^{(k)} & = E[Z_i | p^{(k)}, \mu_1^{(k)}, \mu_2^{(k)}, i, n_i]\\
& = P(Z_i = 1 | p^{(k)}, \mu_1^{(k)}, \mu_2^{(k)}, i, n_i)\\
& = \frac{P(i | Z_i = 1, \mu_1^{(k)})P(Z_i = 1 | p^{(k)})}{P(i | Z_i = 1, \mu_1^{(k)}, \mu_2^{(k)})P(Z_i = 1 | p^{(k)}) + P(i | Z_i = 0, \mu_2^{(k)})P(Z_i = 0 | p^{(k)})} \\
& = \frac{p^{(k)}e^{-\mu_1^{(k)}}(\mu_1^{(k)})^i/i!}{p^{(k)}e^{-\mu_1^{(k)}}(\mu_1^{(k)})^i/i! + (1-p^{(k)})e^{-\mu_2^{(k)}}(\mu_2^{(k)})^i/i!}
\end{align*}

\item M-step

We take the derivative of $Q$ function with respect to $p, \mu_1, \mu_2$ and set to zero in order to derive the estimates of $(k+1)^{th}$ iteration.

Let

\begin{align*}
\frac{dQ}{dp} & = \frac{\sum_i n_ip_i^{(k)}}{p} - \frac{\sum_i n_i(1-p_i^{(k)})}{1-p} = 0 
\end{align*}

So, \begin{align*}
\frac{\sum_i n_i p_i^{(k)}}{\sum_in_i - \sum_in_ip_i^{(k)}} & = \frac{p}{1-p}\\
p^{(k+1)} & = \frac{\sum_i n_i p_i^{(k)}}{\sum_i n_i}.
\end{align*}

Let

\begin{align*}
\frac{dQ}{d\mu_1} & = \sum_i n_i p_i^{(k)}(-1 + \frac{i}{\mu_1}) = 0 
\end{align*}

So, \begin{align*}
\sum_i n_i p_i^{(k)} & = \frac{\sum_i in_ip_i^{(k)}}{\mu_1}\\
\mu_1^{(k+1)} & = \frac{\sum_iin_ip_i^{(k)}}{\sum_in_ip_i^{(k)}}.
\end{align*}

Similarly we can derive that $$\mu_2^{(k+1)} = \frac{\sum_iin_i(1-p_i^{(k)})}{\sum_in_i(1-p_i^{(k)})}.$$
\end{itemize}

\subsection{Factor analysis}\label{facem}
\begin{itemize}

\item E-step

\begin{align*}
& Q(\beta, \tau^2 | \beta^{(k)}, \tau^{(k)}) \\
& = E[ \log{P(Y,Z)} | Y,\beta^{(k)}, \tau^{(k)}) ] \\
& = \sum_i E[ \log{P(Y_i, Z_i)} | Y_i,\beta^{(k)}, \tau^{(k)})] \\
& = \sum_i E[ \log{[P(Y_i | Z_i) P(Z_i) ]} | Y_i,\beta^{(k)}, \tau^{(k)}] \\
& = \sum_i E[ \log{[(2\pi)^{p/2}{|\tau^2|}^{-1/2}\exp{\{-\frac{1}{2}(Y_i - \beta^\top Z_i)^\top {\tau^2}^{-1}(Y_i - \beta^\top Z_i)\}} P(Z_i) ]} | Y_i,\beta^{(k)}, \tau^{(k)})] \\
& = C - \frac{n}{2}\log{|\tau^2|} - \sum_i E[\frac{1}{2}Y_i^\top {\tau^2}^{-1}Y_i - Y_i^\top {\tau^2}^{-1}\beta^\top Z_i + 
\frac{1}{2}Z_i^\top \beta{\tau^2}^{-1}\beta^\top Z_i | Y_i,\beta^{(k)}, \tau^{(k)}] \\
& = C - \frac{n}{2}\log{|\tau^2|} - \sum_i (\frac{1}{2}Y_i^\top {\tau^2}^{-1}Y_i - Y_i^\top {\tau^2}^{-1}\beta^\top E[Z_i|Y_i,\beta^{(k)}, \tau^{(k)}] \\
& \qquad + \frac{1}{2}tr\{ \beta {\tau^2}^{-1} \beta^\top  E[Z_iZ_i^\top |Y_i,\beta^{(k)}, \tau^{(k)}]\}),
\end{align*}
where $C$ is a constant that does not depend on parameters and $k$ denotes the current estimates of parameters. 

Let us define $C_{yz} = \sum_{1}^n \frac{Y_iZ_i^\top }{n}$, $C_{zz} = \sum_{1}^n \frac{Z_iZ_i^\top }{n}$. So the expected value in E-step depends on conditional expectations of the statisitcs $C_{yy}, C_{yz}$ and $C_{zz}$ given observed data matrix $Y$ and current estimates of ${\tau^2}^{(k)}, \beta^{(k)}$. Let $\delta = ({\tau^2}^{(k)} + {\beta^{(k)}}^\top \beta^{(k)})^{-1}({\beta^{(k)}}^\top)$, so by multivariate normal conditional distribution, $E[Z_i|Y_i,\beta^{(k)}, {\tau^2}^{(k)}] = \delta^\top Y_i$. Let $\Delta = Var[Z_i | Y_i,\beta^{(k)}, \tau^{(k)}] = I - \beta^{(k)}({\tau^2}^{(k)}+{\beta^{(k)}}^\top \beta^{(k)})^{-1}{\beta^{(k)}}^\top$.

Therefore, 

\begin{align*}
E[C_{yy} | Y, {\tau^2}^{(k)}, \beta^{(k)}] & = C_{yy} \\
E[C_{yz} | Y, {\tau^2}^{(k)}, \beta^{(k)}] & = \sum_i \frac{E[Y_iZ_i^\top |Y_i,\beta^{(k)}, {\tau^2}^{(k)}]}{n} \\
& =  \sum_i \frac{Y_iY_i^\top }{n} \delta\\
& = C_{yy}\delta \\
E[C_{zz} | Y, {\tau^2}^{(k)}, \beta^{(k)}] & = \sum_i \frac{E[Z_iZ_i^\top |Y_i,\beta^{(k)}, {\tau^2}^{(k)}]}{n}\\
& = \sum_i \frac{Var[Z_i | Y_i,\beta^{(k)}, \tau^{(k)}] + E[Z_i|Y_i,\beta^{(k)}, {\tau^2}^{(k)}]E[Z_i^\top |Y_i,\beta^{(k)}, {\tau^2}^{(k)}]}{n} \\ 
& = \sum_i \frac{\Delta + \delta^\top Y_iY_i^\top \delta}{n}\\
& = \delta^\top  \frac{\sum_i Y_iY_i^\top }{n}\delta + \Delta \\
& = \delta^\top C_{yy}\delta + \Delta.
\end{align*}

\item M-step

If the loading matrix $\beta$ is unrestriced:

In order to obtain the maximizer of $Q$ function in E-step, we take the derivative of $Q$ function with respect to $\beta, {\tau^2}^{-1}$(for convenience) and set to zero. 

\begin{align*}
\frac{dQ(\beta, \tau^2 | \beta^{(k)}, \tau^{(k)})}{d\beta} & = \sum_i {\tau^2}^{-1}Y_iE[Z_i^\top |Y_i,\beta^{(k)}, {\tau^2}^{(k)}] \\
& \qquad - \sum_i {\tau^2}^{-1}\beta^\top E[Z_iZ_i^\top |Y_i,\beta^{(k)}, {\tau^2}^{(k)}] \\
& = 0
\end{align*}

so,
\begin{align*}
\sum_i {\tau^2}^{-1}\beta^\top E[Z_iZ_i^\top |Y_i,\beta^{(k)}, {\tau^2}^{(k)}] & = \sum_i {\tau^2}^{-1}Y_iE[Z_i^\top |Y_i,\beta^{(k)}, {\tau^2}^{(k)}]\\
\beta^\top (\sum_i E[Z_iZ_i^\top |Y_i,\beta^{(k)}, {\tau^2}^{(k)}]) & = \sum_i Y_iE[Z_i^\top |Y_i,\beta^{(k)}, {\tau^2}^{(k)}]\\
\beta^\top E[C_{zz} | Y_i,\beta^{(k)}, {\tau^2}^{(k)}] & = E[C_{yz} | Y_i,\beta^{(k)}, {\tau^2}^{(k)}]\\
\beta^\top (\delta^\top C_{yy}\delta + \Delta) & = C_{yy}\delta\\
\beta^{(k+1)} & = (\delta^\top C_{yy}\delta + \Delta)^{-1}(C_{yy}\delta)^\top 
\end{align*}

\begin{align*}
\frac{dQ(\beta, \tau^2 | \beta^{(k)}, \tau^{(k)})}{d{\tau^2}^{-1}} & = \frac{n}{2}\tau^2 - \sum_i(\frac{1}{2}Y_iY_i^\top  - {\beta^{(k+1)}}^\top E[Z_i|Y_i,\beta^{(k)}, {\tau^2}^{(k)}]Y_i^\top  \\
& \qquad + \frac{1}{2}{\beta^{(k+1)}}^\top E[Z_iZ_i^\top |Y_i,\beta^{(k)}, {\tau^2}^{(k)}]\beta^{(k+1)}) \\
& = 0
\end{align*}

so,
\begin{align*}
\frac{n}{2}\tau^2 & = \sum_i(\frac{1}{2}Y_iY_i^\top  - {\beta^{(k+1)}}^\top E[Z_i|Y_i,\beta^{(k)}, {\tau^2}^{(k)}]Y_i^\top  \\
& \qquad + \frac{1}{2}{\beta^{(k+1)}}^\top E[Z_iZ_i^\top |Y_i,\beta^{(k)}, {\tau^2}^{(k)}]\beta^{(k+1)}) \\
\tau^2 & = \sum_i(\frac{Y_iY_i^\top }{n} - 2{\beta^{(k+1)}}^\top E[\frac{Z_iY_i^\top }{n}|Y_i,\beta^{(k)}, {\tau^2}^{(k)}] \\
& \qquad + {\beta^{(k+1)}}^\top E[\frac{Z_iZ_i^\top }{n}|Y_i,\beta^{(k)}, {\tau^2}^{(k)}]\beta^{(k+1)}) \\
\tau^2 & = C_{yy} - 2{\beta^{(k+1)}}^\top E[C_{yz}^\top |Y_i,\beta^{(k)}, {\tau^2}^{(k)}] \\
& \qquad + {\beta^{(k+1)}}^\top E[C_{zz}|Y_i,\beta^{(k)}, {\tau^2}^{(k)}]\beta^{(k+1)}\\
\tau^2 & = C_{yy} - 2{\beta^{(k+1)}}^\top (C_{yy}\delta)^\top  \\
& \qquad + {\beta^{(k+1)}}^\top (\delta^\top C_{yy}\delta + \Delta)(\delta^\top C_{yy}\delta + \Delta)^{-1}(C_{yy}\delta)^\top \\
\tau^2 & = C_{yy} - 2{\beta^{(k+1)}}^\top (C_{yy}\delta)^\top  + {\beta^{(k+1)}}^\top (C_{yy}\delta)^\top \\
\tau^2 & = C_{yy} - {\beta^{(k+1)}}^\top (C_{yy}\delta)^\top \\
{\tau^2}^{(k+1)} & = \text{diag}\{ C_{yy} - C_{yy}\delta(\delta^\top C_{yy}\delta + \Delta)^{-1}(C_{yy}\delta)^\top \}.
\end{align*}

\end{itemize}

\subsection{Interval censoring}\label{intsem}
\begin{itemize}
\item E-step

$\alpha_{ij}$ tells us the possibility that the event for individual $i$ can occur in interval $(s_{j-1}, s_{j})$ but we do not observe whether it actually occurs. Let us encode this missing information in a new defined variable $Z_{ij}$ such that:
$$Z_{ij} = \begin{cases}
1 & \text{if the event for individual $i$ occurs in $(s_{j-1}, s_{j})$}\\
0 & \text{otherwise}\\
\end{cases}.$$

With this missing variable defined, we now write the $Q$ function in E-step.

\begin{align*}
Q( p | p^{(k)}) & = E[\log{(\prod_i \prod_j p_{j}^{Z_{ij}})} | p^{(k)}, \alpha]\\
& = \sum_i \sum_j \log{p_{j}} E[Z_{ij} | p^{(k)}, \alpha] \\
\end{align*}

Thus $Q$ function depends on the conditional expectation of missing variable $Z_{ij}$ given $\alpha$ matrix and current estimates $p^{(k)}$. By Bayesian formula, 

\begin{align*}
\mu_{ij} & = E[Z_{ij} | p^{(k)}, \alpha] \\
& = P[Z_{ij} = 1 | p^{(k)}, \alpha] \\
& = \frac{\alpha_{ij}p_j}{\sum_{s} \alpha_{is}p_s}, i=1,2,\hdots,n,j=1,2,\hdots,m.
\end{align*}

So, \begin{align*}
Q( p | p^{(k)}) & = \sum_i \sum_j \log{p_{j}} \mu_{ij}
\end{align*}

\item M-step

The probability vector that maximizes $Q$ function serves as the new estimates $p^{(k+1)}$. We introduce a lagrange multiplier $\lambda$ to incorporate the constraint that $\sum_j p_j = 1$. Take derivative with respect to $p_j$,

\begin{align*}
\frac{dQ( p | p^{(k)})}{dp_j} & = \frac{d (\sum_i \sum_j \log{p_{j}} \mu_{ij} + \lambda(1 - \sum_j p_j))}{dp_j} \\
& = \sum_i \frac{\mu_{ij}}{p_j} - \lambda \\ 
& = 0
\end{align*}

So, \begin{align*}
p_j & = \frac{\sum_i \mu_{ij}}{\lambda} \\
\sum_j p_j & = 1 = \frac{\sum_i \sum_j \mu_{ij}}{\lambda} \\
\lambda & = n
\end{align*}

Therefore, \begin{align*}
p_j & = \frac{\sum_i \mu_{ij}}{\lambda}\\
p_j^{(k+1)} & = \frac{1}{n}\sum_i \mu_{ij}, j=1,2,\hdots,m\\
p^{(k+1)} & = (p_1^{(k+1)}, p_2^{(k+1)}, \hdots, p_m^{(k+1)})^\top 
\end{align*}

\end{itemize}

\subsection{Genetics global ancestry estimation problem}\label{genem}
EM Algorithm can be used to compute the maximum likelihood estimates of matrix F and Q. Let us introduce four missing variables for each individual $i$ and marker $j$, $u_{ij}^{(pat)}, u_{ij}^{(mat)},z_{ij}^{(pat)}$ and $z_{ij}^{(mat)}$. $u_{ij}^{(pat)}$ and $u_{ij}^{(mat)}$ represent unobserved allele 1 count from paternal chromosome and maternal chromesome respectively with possible values 0 or 1 while $z_{ij}^{(pat)}$ and $z_{ij}^{(mat)}$ refer to the ancestral populations for the paternal and maternal alleles with possible values $1,\hdots,K$. 

\begin{itemize}
\item E-step

\begin{align*}
& Q(F,Q | F^{(0)}, Q^{(0)}) \\
& = E[\log{\prod_i \prod_j P(x_{ij}, u_{ij}^{(pat)}, u_{ij}^{(mat)},z_{ij}^{(pat)}, z_{ij}^{(mat)} | X, F^{(0)}, Q^{(0)})}]\\
& = E[\log{\prod_i \prod_j (P(x_{ij} | u_{ij}^{(pat)}, u_{ij}^{(mat)}) \prod_k \prod_a(q_{ik} f_{jk}^{I(u_{ij}^{a}=1)}(1-f_{jk})^{I(u_{ij}^{a}=0)})^{I(z_{ij}^{a} = k)})} | X, F^{(0)}, Q^{(0)}]\\
& = \sum_j \sum_k n_{jk}^{(1)}\log{f_{jk}} + n_{jk}^{(0)}\log{(1-f_{jk})} + \sum_i \sum_k m_{ik}\log{q_ik} + C'
\end{align*}

where $C'$  is a constant that does not contain parameters of interest, $$n_{jk}^{(u)} = \sum_i \sum_a P(z_{ij}^{(a)} = k, u_{ij}^{(a)} = u | X, F^{(0)}, Q^{(0)}), u = 0, 1$$ and $$m_{ik} = \sum_j \sum_a P(z_{ij}^{(a)} = k | X, F^{(0)}, Q^{(0)}).$$

To compute $n_{jk}^{(u)}$,$u = 0, 1$ and $m_{ik}$, we need to compute the joint posterior probabilities:
\begin{align*}
& P(u_{ij}^{(pat)}, u_{ij}^{(mat)},z_{ij}^{(pat)}, z_{ij}^{(mat)} | f_{ij}, q_{ik}, x_{ij}) \\
& \propto P(x_{ij} | u_{ij}^{(pat)}, u_{ij}^{(mat)})\prod_k \prod_a(q_{ik} f_{jk}^{I(u_{ij}^{a}=1)} (1-f_{jk})^{I(u_{ij}^{a}=0)})^{I(z_{ij}^{a} = k)} \\
& = I(x_{ij} = u_{ij}^{(pat)} + u_{ij}^{(mat)})\prod_k \prod_a (q_{ik} f_{jk}^{I(u_{ij}^{a}=1)} (1-f_{jk})^{I(u_{ij}^{a}=0)})^{I(z_{ij}^{a} = k)}
\end{align*}

Thus, $n_{jk}^{(u)}$,$u = 0, 1$ and $m_{ik}$ for the $t^{th}$, $t=0,1,\hdots,$ iteration are computed by summing the joint posterior probabilities using $F^{(t)}$ and $Q^{(t)}$.

\item M-step

Take the derivative of $Q$ function with respect to $f_{jk}$ and $q_{ik}$ gives us the next iteration of matrix F and Q that $$f_{jk} = \frac{n_{jk}^{(1)}}{n_{jk}^{(1)} + n_{jk}^{(0)}}$$ and $$q_{ik} = \frac{m_{ik}}{\sum_k m_{ik}}.$$

\end{itemize}

\end{appendices}

\bibliographystyle{natbib}
\bibliography{sq1}

\end{document}